\documentclass[11pt]{article}
\pdfoutput=1
\usepackage[margin=1.0in]{geometry}
\usepackage{amsmath,amssymb,graphicx}
\usepackage{hyperref}
\usepackage{slashed}
\usepackage{physics}

\usepackage[T1]{fontenc}

\usepackage[utf8]{inputenc}

\usepackage[greek,english]{babel}

\usepackage{framed}

\usepackage[font=small,labelfont=bf]{caption}
\usepackage[font=small,labelfont=bf]{subcaption}
\usepackage{cite}

\usepackage{ dsfont }

\usepackage[normalem]{ulem}

\graphicspath{{./Figures/}}

\newcommand{\iu}{{\mathrm i}}
\newcommand{\E}{{\mathrm e}}
\newcommand{\cpi}{\text{\greektext p}}

\newcommand{\RR}{{\mathbb{R}}}
\newcommand{\ZZ}{{\mathbb{Z}}}

\newcommand{\thetabar}{{\overline{\theta}}}

\newcommand{\NDW}{{N_\textsc{DW}}}

\newcommand{\pt}{{\widetilde{p}}}
\newcommand{\qt}{{\widetilde{q}}}

\newcommand{\cL}{\mathcal{L}}

\newcommand{\rmd}{\textrm{d}}

\newcommand{\Uone}{{\mathrm{U(1)}}}
\newcommand{\SU}{{\mathrm{SU}}}
\newcommand{\SUN}{{\mathrm{SU}(n)}}

\newcommand{\Sp}{{\mathrm{Sp}}}
\newcommand{\SO}{{\mathrm{SO}}}

\newcommand{\calS}{{\mathcal{S}}}

\newcommand{\SUNL}{{\mathrm{SU}(n)_\textsc{L}}}
\newcommand{\SUNR}{{\mathrm{SU}(n)_\textsc{R}}}
\newcommand{\SUNV}{{\mathrm{SU}(n)_\textsc{V}}}

\newcommand{\tQ}{{\widetilde{Q}}}

\newcommand{\be}{\begin{equation}}
\newcommand{\ee}{\end{equation}}

\usepackage{xcolor}
\usepackage{bm}

\newcommand{\drawsquare}[2]{\hbox{%
\rule{#2pt}{#1pt}\hskip-#2pt
\rule{#1pt}{#2pt}\hskip-#1pt
\rule[#1pt]{#1pt}{#2pt}}\rule[#1pt]{#2pt}{#2pt}\hskip-#2pt
\rule{#2pt}{#1pt}}

\newcommand{\Yfund}{\raisebox{-.5pt}{\drawsquare{6.5}{0.4}}}
\newcommand{\Ythrees}{\raisebox{-.5pt}{\drawsquare{6.5}{0.4}}\hskip-0.4pt%
          \raisebox{-.5pt}{\drawsquare{6.5}{0.4}}\hskip-0.4pt%
          \raisebox{-.5pt}{\drawsquare{6.5}{0.4}}}
\newcommand{\Yasymm}{\raisebox{-3.5pt}{\drawsquare{6.5}{0.4}}\hskip-6.9pt%
        \raisebox{3pt}{\drawsquare{6.5}{0.4}}}
\newcommand{\Ythreea}{\raisebox{-3.5pt}{\drawsquare{6.5}{0.4}}\hskip-6.9pt%
        \raisebox{3pt}{\drawsquare{6.5}{0.4}}\hskip-6.9pt
        \raisebox{9.5pt}{\drawsquare{6.5}{0.4}}}

\newcommand{\Yadjoint}{\raisebox{-3.5pt}{\drawsquare{6.5}{0.4}}\hskip-6.9pt%
        \raisebox{3pt}{\drawsquare{6.5}{0.4}}\hskip-0.4pt
        \raisebox{3pt}{\drawsquare{6.5}{0.4}}}
\title{\bf  The Quality/Cosmology Tension for a\\ Post-Inflation QCD Axion}
\author{Qianshu Lu,${}^{1,2}$  Matthew Reece,${}^3$ and Zhiquan Sun${}^4$ \\[10pt]
{\small ${}^1$ School of Natural Sciences, Institute for Advanced Study, Princeton, NJ 08540, USA}\\
{\small ${}^2$ Center for Cosmology and Particle Physics, Department of Physics,}\\
{\small New York University, New York, NY 10003, USA}\\
{\small ${}^3$Department of Physics, Harvard University, Cambridge, MA, 02138, USA}\\
{\small ${}^4$Center for Theoretical Physics, Massachusetts Institute of Technology, Cambridge, MA, 02139, USA}\\
}

\begin{document}

\begingroup
{\flushright MIT-CTP 5644\par}
\let\newpage\relax%
\maketitle
\endgroup

\maketitle

\begin{abstract}
It is difficult to construct a post-inflation QCD axion model that solves the axion quality problem (and hence the Strong CP problem) without introducing a cosmological disaster. In a post-inflation axion model, the axion field value is randomized during the Peccei-Quinn phase transition, and axion domain walls form at the QCD phase transition. We emphasize that the gauge equivalence of all minima of the axion potential (i.e., domain wall number equals one) is insufficient to solve the cosmological domain wall problem. The axion string on which a domain wall ends must exist as an individual object (as opposed to a multi-string state), and it must be produced in the early universe. These conditions are often not satisfied in concrete models. Post-inflation axion models also face a potential problem from fractionally charged relics; solving this problem often leads to low-energy Landau poles for Standard Model gauge couplings, reintroducing the quality problem. We study several examples, finding that models that solve the quality problem face cosmological problems, and vice versa. This is not a no-go theorem; nonetheless, we argue that it is much more difficult than generally appreciated to find a viable post-inflation QCD axion model. Successful examples may have a nonstandard cosmological history (e.g., multiple types of cosmic axion strings of different tensions), undermining the widespread expectation that the post-inflation QCD axion scenario predicts a unique mass for axion dark matter.
\end{abstract}

\tableofcontents

\section{Introduction}
\label{sec:intro}

The QCD axion, which dynamically relaxes $\thetabar$ and explains the absence of an observed neutron EDM, has long been the leading contender to solve the Strong CP problem~\cite{Peccei:1977ur, Peccei:1977hh, Weinberg:1977ma, Wilczek:1977pj}. Via the misalignment mechanism, a QCD axion also provides some abundance of dark matter, which could range from a subdominant contribution to one so large it is observationally ruled out, depending on the axion mass and cosmological history~\cite{Preskill:1982cy, Dine:1982ah, Abbott:1982af}. For a detailed introduction to axion physics and the Strong CP problem, see~\cite{Kim:2008hd, Hook:2018dlk, Reece:2023czb}; for axion cosmology specifically, see~\cite{Kawasaki:2013ae, Marsh:2015xka, Safdi:2022xkm}. 

It was appreciated early on that a QCD axion is not truly sufficient to solve the Strong CP problem: the coupling that allows $\thetabar$ to relax explicitly breaks the axion shift symmetry, so a viable theory should explain why there are not other shift-symmetry breaking effects that add to the potential and (generically) displace its minimum. This has come to be known as the ``axion quality problem,'' and it is severe. Even Planck-suppressed higher dimension operators that break a Peccei-Quinn (PQ) symmetry are dangerous, up to dimension ten or more~\cite{Georgi:1981pu, Lazarides:1985bj, Casas:1987bw, Kamionkowski:1992mf, Holman:1992us, Barr:1992qq, Ghigna:1992iv, Randall:1992ut, Dine:1992vx, Kallosh:1995hi}. Depending on the UV structure of the model, PQ symmetry breaking operators are not necessarily Planck-suppressed even when gravity is the only source of PQ symmetry breaking, exacerbating the quality problem~\cite{Contino:2021ayn, Bonnefoy:2022vop}. An axion model that does not solve the axion quality problem should not be thought of as a solution to the Strong CP problem, as it requires introducing many exponentially small parameters rather than than simply assuming $|\thetabar|$ to be small. Solutions to the problem generally either invoke new gauge symmetries that forbid all of the dangerous operators, or extra dimensions so that new terms in the axion potential only arise from nonlocal effects that are exponentially small. 

Axion models are divided into two broad categories based on their cosmology, often referred to as ``pre-inflation'' and ``post-inflation.''\footnote{Alternatively, as an intermediate case, PQ symmetry could have a phase transition during inflation~\cite{Redi:2022llj, Gorghetto:2023vqu}. We will not discuss this case in detail, but for our purposes it may be thought of as similar to either pre-inflation or post-inflation depending on whether or not it introduces a potential domain wall problem in the late-time universe.} In the post-inflation models, there is a phase transition after the end of inflation in which an approximate global PQ symmetry is spontaneously broken. The axion emerges as an independent real scalar field only after this phase transition. In the pre-inflation models, there is no such phase transition. ``Pre-inflation'' is something of a misnomer, because it suggests that there is a PQ symmetry that was realized at some early time in the history of the universe, and this need not be true. For example, models in which the axion arises as a mode of an extra dimensional gauge field (e.g.,~\cite{Witten:1984dg, Choi:2003wr, Svrcek:2006yi, Conlon:2006tq}) offer a compelling solution to the quality problem, and do not have a PQ phase transition at all. They are intrinsically of pre-inflation type (as noted in, e.g.,~\cite{Cicoli:2022fzy,Reece:2023czb, Benabou:2023npn}).

In this paper, we focus instead on the post-inflation axion scenario. In this case, the axion generally arises as the phase of a complex (possibly composite) scalar field, which obtains a vacuum expectation value (VEV) during the PQ phase transition after inflation. This scenario occurs quite generally when the axion decay constant is small compared to the inflationary Hubble scale, $f_a < H_I/(2\cpi)$. It can also arise even for larger $f_a$ in more model-dependent ways, e.g., if the PQ phase transition temperature is much lower than $f_a$, perhaps because it is triggered by SUSY breaking~\cite{Asaka:1998ns, Banks:2002sd}; or if a sufficiently high temperature is achieved during reheating to restore the PQ symmetry~\cite{Hertzberg:2008wr}; or if the heavy lifting effect raises the mass of the PQ scalar during inflation~\cite{Bao:2022hsg}.

The post-inflation scenario has a rich cosmology. During the PQ phase transition, the value of the axion field becomes randomized in different parts of the universe. Axion strings, around which the axion field value winds, are produced through the Kibble-Zurek mechanism. Later, during the QCD phase transition, the axion acquires a potential and domain walls form that end on the axion strings. Axion dark matter arises not only through the misalignment mechanism, but from radiation from the strings after the PQ phase transition and from the dynamics of the string-domain wall network after the QCD phase transition. 

Domain walls are potentially disastrous for the cosmology of a post-inflation axion~\cite{Sikivie:1982qv}, because their energy density redshifts more slowly than that of matter or radiation and can come to dominate the universe~\cite{Zeldovich:1974uw}. There are two mechanisms by which cosmological domain walls can collapse: they can either {\em have} a boundary, ending on a cosmic string\cite{Kibble:1982ae, Vilenkin:1982ks, Kibble:1982dd, Everett:1982nm}, or they can {\em be} the boundary of a volume of higher vacuum energy, which exerts a force to collapse the wall. The former case occurs when the broken discrete symmetry is gauged, and the latter case occurs when the discrete symmetry is explicitly broken. Explicit symmetry breaking has been proposed as a solution to the axion domain wall problem~\cite{Sikivie:1982qv, Gelmini:1988sf, Larsson:1996sp, Hiramatsu:2012sc}, but it has an obvious tension with the quality problem. PQ-violating terms in the potential must be large enough to cause domain walls to collapse quickly, but small enough to maintain $|\thetabar| \lesssim 10^{-10}$. A related proposal to solve the problem is to begin with biased initial conditions, which again can cause domain walls to collapse~\cite{Coulson:1995nv,Hindmarsh:1996xv,Larsson:1996sp}. This requires either explicit symmetry violation at earlier times (again in obvious tension with the quality problem) or an unconventional cosmology. The tension between the quality problem and a fast enough decay of the domain wall in this scenario is long known and well-studied in the literature. The decay of domain walls due to explicit PQ violation in the potential copiously produces axion dark matter. Quantifying this effect has been the aim of studies based on detailed numerical simulations and scaling arguments~\cite{Hiramatsu:2010yn, Hiramatsu:2012sc, Kawasaki:2014sqa, Beyer:2022ywc, Chang:2023rll}. These papers argue that if the CP phase of the PQ-violating term is $O(1)$, the current bound $|\thetabar|\lesssim 10^{-10}$ and the constraint that axion dark matter is not overproduced completely eliminate the parameter space with $f_a \gtrsim 10^8\,\mathrm{GeV}$ that is consistent with astrophysical bounds on the QCD axion. Only by tuning the CP phase to be small (and thus not fully solving the quality problem) can this scenario have a consistent cosmology. We will not discuss explicit breaking or biased initial conditions further in this paper, and focus instead on the alternative scenario of gauged symmetry, where the tension between the quality problem and cosmology has not been clarified in previous work.

In this work, we assume the simplest solution to the axion domain wall problem: destruction of the axions by a network of cosmic strings that formed during the PQ phase transition. This mechanism works when a single domain wall can end on a cosmic string~\cite{Lazarides:1982tw}. The number of axion domain walls ending on an axion string of minimal winding number, often simply called the ``domain wall number,'' can be read off from the axion-gluon coupling. Specifically, given a $2\cpi$-periodic axion field $\theta(x)$ coupling to gluons via the action
\begin{equation}
\frac{k_G}{8\cpi^2} \int \theta(x)\, \mathrm{tr}(G \wedge G),
\end{equation}
we have $k_G \in \ZZ$ for topological reasons, and QCD dynamics generates a potential with the property $V(\theta) = V(\theta + 2\cpi/k_G)$. Thus, there are $|k_G|$ degenerate minima, and we say the theory has domain wall number $\NDW = |k_G|$. Hence, the simplest solution to the axion domain wall problem is to engineer a model with $\NDW = 1$, and to ensure that axion strings of minimal winding number are produced during the PQ phase transition.

The post-inflation axion scenario with $\NDW = 1$ is often, implicitly or explicitly, assumed in discussions of axion phenomenology and the detection of axion dark matter. In the pre-inflation scenario, a wide range of axion masses can accommodate the observed dark matter abundance, depending on initial conditions. The post-inflation axion is often perceived as more predictive. To quote some recent statements in the literature: ``there is in principle a unique calculable prediction for the axion mass if it is to make up the complete DM density in such [post-inflation] models, which would be extremely valuable for experimental axion searches''~\cite{Gorghetto:2018myk}; ``If the PQ symmetry is broken after the cosmological epoch of inflation, then there is a unique axion mass $m_a$ that leads to the observed DM abundance''~\cite{Buschmann:2021sdq}. Sophisticated numerical simulations have been undertaken to assess the dark matter abundance in a post-inflation QCD axion scenario~\cite{Buschmann:2019icd, Hiramatsu:2012gg, Klaer:2017ond, Gorghetto:2018myk, Gorghetto:2020qws, Buschmann:2021sdq}, and to predict a specific mass for which axions constitute all of the dark matter. These computations show that axion strings formed at the PQ phase transition enter a scaling regime, emitting QCD axion dark matter as the string network evolves, and then the string-domain wall network tears itself apart after the QCD phase transition. The axions emitted from string and domain wall dynamics add to the abundance of dark matter arising from misalignment. Axion experimentalists take the predictions of such simulations seriously for determining the optimal mass range to target in their searches.

In this paper, we complicate this picture. We argue that there is a tension between achieving $\NDW = 1$ and solving the axion quality problem without introducing other disastrous cosmological problems. We illustrate this with various examples. While we do not have a general no-go theorem, we also do not know any example of a QCD axion model without a quality problem that gives rise to the conventional post-inflation axion cosmology. If there is an axion theory that solves the quality problem and is not cosmologically excluded, it seems likely that it also has a non-standard cosmological history, which would alter the prediction for the axion dark matter abundance. One message to take away is that experiments searching over a broad mass range, rather than targeting any theorists' preference, are vital. Another is that axion cosmology could differ in a variety of ways from that studied in existing simulations; for example, there could be a network consisting of multiple types of axion strings with parametrically different tensions at the time of the QCD phase transition. Given the computational resources that are now being deployed to study post-inflation axion cosmology, it would be worthwhile to carry out a suite of simulations for a wider range of cosmological scenarios.

This paper is organized as follows. 
In Sec.~\ref{sec:Zn}, we discuss the canonical solution to the axion quality problem: a discrete $\ZZ_p$ symmetry. We point out a basic cosmological tension. The $\ZZ_p$ symmetry ensures that, at the time of the QCD phase transition, domain walls can form separating at least $p$ different vacua. When $\ZZ_p$ is a gauge symmetry, these vacua are gauge equivalent and the domain wall can, in principle, end. However, whether the domain wall can end on a single string, and whether cosmological dynamics produces such strings, depend on details of the UV completion. We discuss a specific UV completion in recent literature where the $\ZZ_p$ symmetry is embedded in the center of a continuous gauge group $\SU(N)$ \cite{Ardu:2020qmo}.
In Sec.~\ref{sec:Uone}, we examine a KSVZ-type model with a $\Uone$ gauge symmetry to ensure a high-quality axion and avoid the domain wall problem \cite{Barr:1992qq}. We find that the model suffers from a series of cosmological problems including fractionally charged relics and Landau poles in the Standard Model couplings below the UV scale. 
In Sec.~\ref{sec:SUnSUn}, we briefly consider a model where the axion quality is protected by a nonabelian gauge symmetry $\SU(n) \times \SU(n)$~\cite{DiLuzio:2017tjx}. The cosmological considerations are similar to the previous section and we arrive at qualitatively the same conclusion.
In Sec.~\ref{sec:composite}, we comment on the class of composite axion models, in which we find it is generally hard to avoid the domain wall problem.
We present discussions on other types of models in the literature in Sec.~\ref{sec:discussion}, and pose some open questions on a more holistic approach to the solution of the axion quality problem, the domain wall problem, and the cosmological dynamics.

\section{A $\ZZ_p$ symmetry for quality}
\label{sec:Zn}

\subsection{First look}
\label{sec:Zn1}

It is often said that the simplest solution to the axion quality problem is to impose a $\ZZ_p$ discrete symmetry, as first studied in~\cite{Lazarides:1985bj}. For example, if a PQ-charged scalar field $\Phi$ carries charge 1 under such a symmetry, then all operators of the form $\Phi^k / \Lambda^{k-4}$ are forbidden up to $k = p$. More generally, we will consider $\Phi$ to carry charge $q$ under the $\ZZ_p$ symmetry. Then we define
\begin{equation}
\label{eq:gcd}
    d \equiv \gcd(q, p), \qquad \pt \equiv \frac{p}{d}.
\end{equation}
In this case, $\Phi^k$ terms are forbidden up to $k = \pt$. Although $q$ is only defined modulo $p$, $d$ and $\pt$ are well-defined integers. When $\Phi$ obtains a vacuum expectation value, $\ZZ_p$ is spontaneously broken to a residual  $\ZZ_d$ symmetry that does not act on the axion.

There is a basic tension between such a solution to the quality problem and the domain wall problem. If the axion arises as the phase of $\Phi$, 
\begin{equation}
\Phi(x) = \frac{1}{\sqrt{2}} \left(f + s(x)\right) \E^{\iu \varphi(x)},
\end{equation} 
then under the generator $z$ of the $\ZZ_p$ symmetry $\Phi \mapsto \E^{2 \cpi \iu q/p}\Phi$ and hence
\begin{equation} \label{eq:fractionalshift}
z: \qquad \varphi(x) \mapsto \varphi(x) + \frac{2\cpi q}{p}.
\end{equation}
(This makes sense as a $\ZZ_p$ symmetry action even though $q$ is only defined mod $p$ because, through its definition as a phase, $\varphi \cong \varphi + 2\cpi$.) The symmetry~\eqref{eq:fractionalshift} allows terms in the potential of the form $\cos(\pt (\varphi - \varphi_0))$, and hence $\pt$ vacua separated by domain walls. For an exact $\ZZ_p$ global symmetry, these are stable domain walls. However, we don't expect global symmetries to ever be exact. For a principled mechanism forbidding higher-dimension operators, we should focus on gauge symmetries.

It is instructive to see how the domain wall number divisible by $\pt$ arises in a detailed model with a $\ZZ_p$ gauge symmetry. For concreteness, consider a KSVZ-type model with quarks $Q^{(i)}, \tQ^{(i)}$ carrying $\ZZ_p$ gauge charges $k, -(k+q)$ respectively, with flavor index $i$ spanning a range $i \in \{1, \ldots, r\}$. With this charge assignment, Yukawa terms
\begin{equation} \label{eq:ZpYukawas}
\sum_{i,j} \left(y_{ij} \Phi Q^{(i)} \tQ^{(j)} + \mathrm{h.c.}\right)
\end{equation}
are allowed by gauge invariance. (In later such expressions, we will leave the sum over $i$ and $j$ implicit.) Such a model potentially has a mixed $\ZZ_p\text{-}\SU(3)_\textsc{C}^2$ anomaly. In particular, under a rephasing
\begin{equation}
\Phi \mapsto \E^{\iu q \alpha} \Phi, \quad Q^{(i)} \mapsto \E^{\iu k \alpha} Q^{(i)}, \quad \tQ^{(i)} \mapsto \E^{-\iu (k+q) \alpha} \tQ^{(i)},
\end{equation}
the chiral anomaly shifts the action by 
\begin{equation}
\Delta I = \frac{1}{8\cpi^2} r \left[k \alpha + (-k - q)\alpha\right] \int \mathrm{tr}(G \wedge G) = -\frac{rq}{8\cpi^2} \alpha \int \mathrm{tr}(G \wedge G).
\end{equation}
In order for the $\ZZ_p$ transformation to be a valid gauge symmetry, we require that $\exp[\iu \Delta I] = 1$ when $\alpha = 2\cpi/p$. The quantization of instanton number $\frac{1}{8\cpi^2} \int \mathrm{tr}(G \wedge G) \in \ZZ$ then implies that
\begin{equation} \label{eq:sinZ}
\frac{rq}{p} \in \mathbb{Z},
\end{equation}
which in turn implies that the number of KSVZ quarks $r$ must be an integer multiple of $\pt$. We denote this integer as $s$:
\begin{equation}
    s \equiv \frac{r}{\pt} \in \ZZ.
\end{equation}
Integrating out the heavy modes after $\Phi$ gets a VEV, we obtain an effective coupling
\begin{equation} \label{eq:varphieffcoupling}
\frac{r}{8\cpi^2} \int \varphi(x) \mathrm{tr}(G \wedge G),
\end{equation}
so the domain wall number $|r|$ associated with $\varphi$ is a multiple of $\pt$.

\subsection{Second look: $\ZZ_p$ strings}
\label{sec:Zn2}

The argument that we have just given is not quite right, although it hints at an important physical effect. Once we gauge $\ZZ_p$, the identification~\eqref{eq:fractionalshift} is a gauge transformation. If all of the vacua separated by domain walls are gauge equivalent under this transformation, then the domain wall problem is potentially solved~\cite{Lazarides:1982tw}. A minimal domain wall interpolates between a minimum at some value $\varphi_0$ and a neighboring minimum at $\varphi_0 + 2\cpi/r$. These two minima are gauge equivalent under $z^k \in \ZZ_p$ if
\begin{equation} \label{eq:equivcond}
    \frac{2\cpi}{r} \equiv \frac{2\cpi k q}{p} \mod 2\cpi \quad \Rightarrow \quad \frac{2\cpi}{s \pt} \equiv \frac{2\cpi k q}{d \pt} \mod 2\cpi.
\end{equation}
But (for any choice of integer $q$ in its equivalence class) $kq/d$ is an integer, which means that in order for this expression to hold, $1/s$ must also be an integer. This is only possible for the special cases $s = \pm 1$; suppose this is true. Choosing a representative integer $q$ from its $\ZZ_p$ equivalence class, we can write $q = \qt d$. The condition~\eqref{eq:equivcond} is then equivalent to 
\begin{equation} \label{eq:equivcond2}
k \qt \equiv \pm 1 \mod \pt. 
\end{equation}
It is a standard result in modular arithmetic that a solution $k$ exists precisely when $\pt$ and $\qt$ are relatively prime, which is the case here (by construction). Hence, we conclude that neighboring minima are gauge equivalent if and only if $|s| = \pm 1$. Indeed, in general, the $\ZZ_p$ orbit generated by~\eqref{eq:fractionalshift} includes $\pt$ distinct gauge equivalent values of $\varphi$, and hence $|s| = |r|/\pt$ is the number of distinct gauge orbits of minima. Thus, $|s|$ is the correct, physical domain wall number in this theory.

We can define a $2\cpi$-periodic axion field $\theta(x)$ by
\begin{equation}
\theta(x) = \pt \varphi(x) = \arg(\Phi(x)^\pt).
\end{equation}
In other words, $\theta$ is the phase of the smallest $\ZZ_p$-invariant operator constructed from $\Phi$. We will see a similar pattern in other examples: the $2\cpi$-periodic axion is often the phase of the lowest-dimension gauge invariant operator that carries PQ charge and obtains a vacuum expectation value.
In terms of this variable, the effective coupling~\eqref{eq:varphieffcoupling} is 
\begin{equation} \label{eq:thetanormcoupling}
\frac{r}{\pt} \frac{1}{8\cpi^2} \int \theta(x) \mathrm{tr}(G \wedge G) = \frac{s}{8\cpi^2} \int \theta(x) \mathrm{tr}(G \wedge G),
\end{equation}
which makes manifest that the true domain wall number is $s$, which could be $\pm 1$. Thus, a priori, a model with a $\ZZ_p$ gauge symmetry protecting axion quality need not be in conflict with a simple solution to the axion domain wall problem.

\begin{figure}
    \centering
    \includegraphics[width=0.5\textwidth]{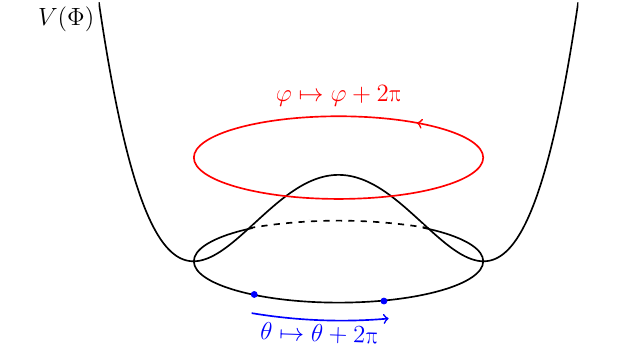}
    \caption{A generic PQ-breaking potential for a complex scalar $\Phi$ with charge $q$ under a gauged $\ZZ_p$ subgroup of $\Uone_\textsc{PQ}$. As in the text, we define $d \equiv \gcd(q, p), \pt \equiv p/d$. The field $\theta = \varphi/\pt$ is $2\cpi$ periodic, where $\varphi$ is the phase of $\Phi$. In the illustration, $\pt = 5$. However, at the time of the PQ phase transition when $\Phi$ obtains a VEV, there is no dynamical reason for the phase of $\Phi$ to be restricted in the sub-interval where $\theta\mapsto \theta + 2\cpi$. The cosmic strings produced therefore wind around the whole $2\cpi$ period of $\varphi$, which will be connected to $\pt$ domain walls after the QCD phase transition.}
    \label{fig:PQ_potential}
\end{figure}
The trouble is that this argument is kinematic: the symmetries of the problem allow for the right states to exist to solve the domain wall problem. They do not, however, guarantee that the {\em dynamics} of the theory will solve the domain wall problem. In particular, at the PQ phase transition, the complex field $\Phi(x)$ gets a VEV from some potential $V(|\Phi|)$. The Kibble-Zurek mechanism guarantees that this phase transition will form axion strings around which the phase $\varphi$ winds from $0$ to $2\cpi$, as illustrated in Fig.~\ref{fig:PQ_potential}.  Equivalently, these are strings for which $\theta(x)$ winds from $0$ to $2\cpi \pt$. Such a string can form a junction on which $\pt$ minimal domain walls end, but it cannot destroy a single domain wall. In other words, {\em as far as the PQ phase transition is concerned}, we have a theory in which the domain wall number is divisible by $\pt$.

The $\ZZ_p$ gauge symmetry allows for the existence of {\em twist vortices}, dynamical cosmic strings around which fields come back to themselves only up to a $\ZZ_p$ gauge transformation~\cite{Krauss:1988zc}. These objects are expected to exist in any theory with a $\ZZ_p$ gauge symmetry, due to the absence of global symmetries in quantum gravity~\cite{Heidenreich:2021xpr} (this is one example of the completeness of the spectrum of quantum gravity~\cite{Polchinski:2003bq}). The existence of these objects, in a theory with $\NDW = 1$, renders domain walls unstable: there is always some stringlike object (possibly a composite one) on which a single domain wall can end. Even if these strings are not populated in the early universe, a domain wall can {\em in principle} decay through the nucleation of a bubble of string within the wall, though this rate is exponentially suppressed when the string tension is much larger than the domain wall tension~\cite{Kibble:1982dd}, so it does not solve the QCD axion's domain wall problem. Instead, we must rely on the cosmological production of appropriate strings for the wall to end on.

In order for $\ZZ_p$ strings to solve the domain wall problem, two important conditions should be satisfied: there should be a single string on which a domain wall can end, and these strings should be produced in the early universe. Neither is guaranteed. For example, consider a minimal domain wall interpolating between $\varphi = 0$ and $\varphi = 2\cpi/\pt$. These vacua are equivalent under the gauge transformation by $z^k$ with $k$ obeying the condition~\eqref{eq:equivcond2}. A $\ZZ_p$ string of charge $k$ has the property that, as we circle the string, the field $\Phi$ maps back to itself up to a phase $\exp(2\cpi \iu k q/p)$. This means that the {\em winding} of $\varphi$ around the $z^k$ string is given by
\begin{equation} \label{eq:winding}
   \oint_{z^k} \rmd \varphi = 2\cpi \left(\frac{k q}{p} + n\right),
\end{equation}
for some integer $n$. There is some choice of $n$ for which this is equal to the desired $2\cpi/\pt$, but whether a string with appropriate $n$ exists depends on the UV completion of the theory. 

\begin{figure}
    \centering
    \begin{subfigure}[b]{0.55\textwidth}
         \centering
         \includegraphics[width=\textwidth]{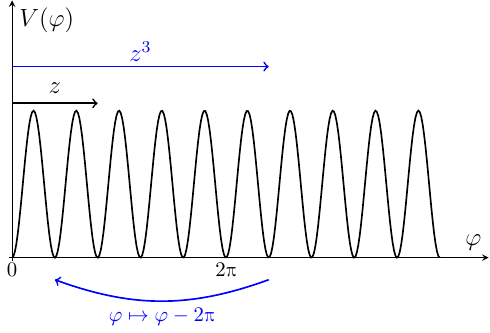}
         \caption{}
         \label{fig:2Z5_potential}
     \end{subfigure}
    \begin{subfigure}[b]{0.35\textwidth}
         \centering
         \includegraphics[width=\textwidth]{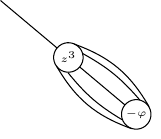}\caption{}
         \label{fig:2Z5_blob}
     \end{subfigure}
    \caption{\textbf{(a)} The potential for $\varphi$ from the phase of $\Phi$, where $\Phi$ has charge 2 under a $\ZZ_5$ gauge symmetry. The action of the $\ZZ_5$ generator $z$ does not map between neighboring vacua in $V(\varphi)$, but the combination of $z^3$ and $\varphi \mapsto \varphi - 2\cpi$ gauge transformation does. This means that the object where a single axion domain wall can end on is the bound state of a $z^3$ string and a ``$-\varphi$'' string (where $\varphi$ winds by $-2\cpi$). In isolation, a $z^3$ string will have six domain walls attached to it, five of which can connect to the same $-\varphi$ string to end the remaining domain wall. Another state on which a single axion domain wall can end is the bound state of a $z^2$ string and a $-\varphi$ string: a $-\varphi$ string is attached to five domain walls, and four of the five domain walls are absorbed by the $z^2$ string.}
    \label{fig:charge2Z5}
\end{figure}

To make this more clear, we consider an example: a theory with $\ZZ_5$ gauge symmetry under which $\Phi$ has charge 2, and we suppose that $s = 1$. The $\ZZ_5$ generator $z$ maps $\varphi \mapsto \varphi + 4\cpi/5 \mod 2\cpi$. Consider the minimal domain wall interpolating between $\varphi_0$ and $\varphi_0 + 2\cpi/5$. Modulo $2\cpi$, these are related not by the generator $z$ itself, but by $z^3$, because $3 \times 2 = 1 + 5 \cong 1 \mod 5$. In~\eqref{eq:winding}, then, this example has $p = 5$, $q = 2$, $k = 3$, and we need $n = -1$ to obtain net winding $\oint_{z^3} \rmd \varphi = 2\cpi\left(3 \times 2/5 - 1\right) = 2\cpi/5$. There may be a model where a string producing such winding exists as a single object, perhaps even an elementary one. (Because $z^3$ also generates $\ZZ_5$, there is no invariant meaning to our choice of $z$ as a generator, and no reason to favor the case where the strings with holonomy $z$ are fundamental.) On the other hand, it may be that the only elementary $\ZZ_5$ string implements the operation $z$ such that the winding is $\oint_z \rmd \varphi = 4\cpi/5$. In this case, the $z^3$ string could arise by stacking three such $z$ strings on top of each other, generating winding $\oint_{z^3} \rmd \varphi = 12\cpi/5$. This overshoots the desired $2\cpi/5$ winding, and we require an additional shift in $\varphi$ back $2\cpi$, as shown in Fig.~\ref{fig:2Z5_potential}. In this case, the desired string on which a single axion domain wall can end is the bound state of a $z^3$ string and (in an abuse of notation) a $-\varphi$ string where $\varphi$ winds by $-2\cpi$. A $z^3$ string has six domain walls attached to it, and five of them are absorbed by the $-\varphi$ string, as illustrated in Fig.~\ref{fig:2Z5_blob}.\footnote{Alternatively, because $z^2$ is the inverse element of $z^3$ in $\ZZ_5$, one could stack two anti-$z$ strings, leading to net winding $-8\cpi/5$, and one $+\varphi$ string of winding $2\cpi$ for a net $2\cpi/5$. A similar picture results.} While $-\varphi$ strings will be produced by the PQ phase transition following the familiar story of cosmic string production from a spontaneously broken global $\Uone$ symmetry, the expected abundance for $z^3$ strings can only be determined within a UV completion. Fig.~\ref{fig:2Z5_blob} is consistent with the general claim that the domain wall can end when $\NDW = 1$. However, the messy, composite configuration on which it ends is unlikely to arise dynamically. More likely, if there is a large abundance of $z^3$ strings and $\varphi$ strings, one would form a large, frustrated network and the domain wall network would not collapse. This is because, effectively, the picture is much like that for $\NDW > 1$: any given elementary string is attached to multiple domain walls. In summary, the cosmological consequences of a discrete gauge symmetry depend on details of the UV completion that cannot be deduced from the symmetry structure alone.

For the moment, suppose that the requisite $\ZZ_p$ strings to destroy minimal domain walls {\em are} somehow produced cosmologically, so that at the time of the QCD confining transition, the universe is filled with a network of them. Axion domain walls can end on these strings, solving the domain wall problem. However, there is an important physical difference from the standard scenario. Because the $\ZZ_p$ strings are not produced in the PQ phase transition, there is no reason for their tension $\mu$ to be tied to the scale $f_a$ of the axion decay constant. In principle, they could have a much larger tension. The string network radiates axions continuously as it evolves, and further emits axions when the string-domain wall network fragments, all of which enhance the abundance of axion dark matter. We will follow a recent discussion in~\cite{Buschmann:2021sdq}. The rate of axion emission from strings is $\Gamma_a \approx 2 H \rho_s$, with $\rho_s$ the energy density in axion strings. This is directly proportional to the string tension: $\rho_s = \xi \mu/t^2$, where $\xi$ is the string length per Hubble volume and is expected to be only logarithmically sensitive to the string tension or other detailed physics in the string core. The final axion abundance is largely determined by this process and proportional to $\Gamma_a$. The estimate of~\cite{Buschmann:2021sdq} assumes a string tension $\mu_0 = \cpi f_a^2$. If the $\ZZ_p$ string tension $\mu$ is much larger, then a first estimate is that the axion dark matter abundance is enhanced by a factor of $\mu/\mu_0$. (A similar discussion of the role of higher tension strings in enhancing the axion emission rate appeared in~\cite{Niu:2023khv}, as this paper was being completed.) This, correspondingly, implies that the correct axion dark matter abundance is attained for a smaller value of $f_a$ and a larger axion mass than in the conventional axion cosmology. We emphasize that this is just a crude scaling argument, and the additional physics needed to construct a model in which the requisite population of $\ZZ_p$ strings is produced cosmologically could further complicate the story. 

Before moving on, let us make a brief comment about Peccei-Quinn charges. In the above discussion, we have never explicitly referred to a $\Uone_\textsc{PQ}$ global symmetry. We could do so as follows: the couplings~\eqref{eq:ZpYukawas} respect a chiral symmetry under which $Q^{(i)}$ has charge $p_Q = +1$, $\Phi$ has charge $p_\Phi = -1$, and $\tQ^{(j)}$ has charge $p_\tQ = 0$. (They also respect a ``heavy baryon number'' symmetry under which $Q^{(i)}$ has charge $+1$ and $\tQ^{(j)}$ has charge $-1$; any linear combination of this and the aforementioned chiral symmetry could be viewed as a Peccei-Quinn symmetry.) A standard formula is that the axion-gluon coupling is given by $k_G = \sum_i 2I(r_i) p_i$, where the sum is over left-handed Weyl fermions with PQ charge $p_i$ and $\SU(3)_\textsc{C}$ representation $r_i$. In our example, this computation gives $k_G = r$, as in~\eqref{eq:varphieffcoupling}. In order to compute the properly normalized coupling $k_G = r/\pt = s$ as in~\eqref{eq:thetanormcoupling}, then, we should normalize the PQ charge so that $p_\Phi = -1/\pt$, i.e., it is the operator $\Phi^\pt$ that has unit magnitude PQ charge. This may seem unusual, since we customarily normalize $\Uone$ charges to be integers so that $\E^{\iu \alpha} \in \Uone$ has a well-defined action on a field as $\psi \mapsto \E^{\iu q \alpha} \psi$. However, fractional normalizations of charges under a global $\Uone$ symmetry are often a useful convention in the case where a discrete subgroup of $\Uone$ is gauged. A familiar example is baryon number symmetry, where we normalize quark fields to have baryon number $1/3$, the $\ZZ_3$ center of $\SU(3)_\textsc{C}$ acts on the quarks in the same manner as the $\ZZ_3$ subgroup of baryon number, and the gauge invariant baryon operators have baryon number 1. The utility of fractional PQ charge assignments will become important in Sec.~\ref{sec:composite} below.

\subsection{$\ZZ_p$ embedded in continuous gauge groups}
\label{sec:Zn3}
We consider an illustrative class of models in a recent work \cite{Ardu:2020qmo}, which aims to produce the required $\ZZ_p$ string at the time of the PQ phase transition. The string tension $\mu$ is then tied with $f_a$, and the model regains the possibility of a precise prediction of the axion mass to produce the correct dark matter abundance. In this class of models, the axion arises from a scalar $\calS$ spontaneously breaking the gauge group $\SU(N)$. $\calS$ lives in either the symmetric or antisymmetric two-index representation of $\SU(N)$, which lead to different types of models as we will discuss shortly. These models are realizations of the Lazarides-Shafi idea first established in \cite{Lazarides:1982tw}, where the domain wall problem is solved by embedding the discrete $\mathbb{Z}_p$ subgroup of $\Uone_\mathrm{PQ}$ in a continuous gauge group.  However, upon closer examination of the models, we find that some subtleties have been overlooked and the embedding is not necessarily sufficient to solve the domain wall problem.

The fermion content and charge assignments of the models are universal regardless of whether $\calS$ is a symmetric or antisymmetric tensor. The fermions include $\SU(N)$ fundamental quarks $Q$ (in the $\bm{3}$ of $\SU(3)_\textsc{C}$) and $\tQ$ (in the $\bm{\overline{3}}$) which we can take to have hypercharge $Y = \mp 1/3$, and $\SU(N)$ antifundamental leptons $L^{(i)}$ and $\widetilde{L}^{(i)}$ ($i = 1,2,3$), which we can take to have $Y = 0$. These have Yukawa interactions with $\calS$,
\be
\mathcal{L}\supset - y_Q Q\calS^* \tQ - \left(y_L\right)_{ij} L^{(i)}\calS\widetilde{L}^{(j)} + \mathrm{h.c.}
\ee
The potential for $\calS$ is constructed such that it obtains a VEV
\be 
\expval{\calS} = v_{\calS}\left(\mathds{1}_{N/2\times N/2}\otimes \epsilon \right)
\ee
when $\calS$ is in the antisymmetric two-index representation and $N$ is even, which spontaneously breaks $\SU(N)$ to $\Sp(N),$\footnote{When $N$ is odd the symmetry breaking pattern is different and there is no light axion~\cite{Buttazzo:2019mvl}.} and
\be \label{eq:Sexpsym}
\expval{\calS} = v_{\calS}\mathds{1}_{N\times N}
\ee
when $\calS$ is in the symmetric two-index representation, which spontaneously breaks $\SU(N)$ to $\SO(N)$. In both cases, we will denote the overall phase of $\calS$ by $\varphi(x)$; that is, we take
\begin{equation}
    \varphi(x) = \frac{1}{N} \arg \det \calS.
\end{equation}
Performing a chiral rotation on the heavy quarks to remove $\varphi(x)$ from the Yukawa terms, we obtain the $\varphi(x)$ coupling to gluons
\be
\mathcal{L}\supset \frac{N}{8\cpi^2}\int \varphi(x)\mathrm{tr}(G \wedge G).
\ee
The domain wall number with respect to a $2\cpi$ period of $\varphi(x)$ is thus $N$, i.e., neighboring vacua in the $\varphi(x)$ potential are separated by $2\cpi/N$.

From the discussion in Sec.~\ref{sec:Zn1}, we know that this is not the whole story, since the $N$ vacua could be gauge equivalent. To see if that is the case, we need to look at the lowest-dimensional $\Uone_\mathrm{PQ}$-charged operator that is consistent with the gauge symmetry of the model. Now we will look at the antisymmetric and symmetric representations of $\calS$ separately. When $\calS$ is in the antisymmetric representation and $N$ is even, the lowest-dimensional operator is the Pfaffian, $\mathrm{Pf}\,\calS = \sqrt{\det\calS} = |\sqrt{\det\calS}|\E^{\iu N\varphi(x)/2}$. The minimal $2\cpi$-invariant variable with no gauge-equivalent sub-interval is therefore $\theta(x) = N\varphi(x)/2$, which we will identify as the axion in our theory. By definition, all cosmic strings must wind by an integer multiple of $2\cpi$ in $\theta(x)$. Since
\be
\mathcal{L}\supset \frac{N}{8\cpi^2}\int \varphi(x)\mathrm{tr}(G \wedge G) = \frac{2}{8\cpi^2}\int \theta(x)\mathrm{tr}(G\wedge G),
\ee
we see that the model has a domain wall number of 2, thus the domain wall problem is not solved. The issue is that even though there is an anomaly-free $\ZZ_N$ inside $\Uone_\textsc{PQ}$, only the $\ZZ_{N/2}$ subgroup is gauged via its embedding in the center $\ZZ_N \subset \SU(N)$ by the model construction.

When $\calS$ is in the symmetric representation, the lowest-dimensional PQ-breaking operator is the determinant of $\calS$, $\det\calS = |\det\calS|\E^{\iu N \varphi(x)}$, for both even and odd $N$. It is tempting to say then that the minimal $2\cpi$ periodic variable is $\theta(x) = N\varphi(x)$. However, when $N$ is even, this is not correct: $\calS$ being a two-index symmetric representation of $\SU(N)$, it has charge 2 under the $\ZZ_N$ center of $\SU(N)$ and is invariant under the $\ZZ_2$ subgroup of $\ZZ_N$. More specifically, under a $z^k\in \ZZ_N$ transformation, $\varphi(x)\mapsto \varphi(x) + 4\cpi k/N$, which cannot relate $\varphi(x)$ with the neighboring vacua $\varphi(x) \pm 2\cpi/N$ when $N$ is even. The minimal $2\cpi$ periodic variable in the model is thus $\theta(x) = N\varphi(x)/2$. Similar to the antisymmetric case, we have
\be
\mathcal{L}\supset \frac{N}{8\cpi^2}\int \varphi(x)\mathrm{tr}(G \wedge G) = \frac{2}{8\cpi^2}\int \theta(x)\mathrm{tr}(G\wedge G),
\ee
and the domain wall number of the model is again $\NDW = 2$.

When $N$ is odd, there is no $\ZZ_2$ subgroup of $\ZZ_N$, and a $z^k\in \ZZ_N$ transformation will be able to reach all vacua starting from $\varphi(x)$. In this case, we take $\theta(x) = N\varphi(x)$, and we have $\NDW = k_G = 1$. All domain walls are unstable in this theory. However, referring back to our example of a $\ZZ_5$ symmetry in Sec.~\ref{sec:Zn2}, the minimal string for the $\ZZ_N$ center cannot destroy a single domain wall, because the minimal string winds $\varphi$ from 0 to $4\cpi/N$ rather than $2\cpi / N$. The strings that may play a dynamical role in destroying domain walls are \emph{non-minimal}, as they are composite objects of $(k+1)$ minimal $z$ strings (or equivalently, $k$ minimal anti-strings), where $N=2k+1$, together with $\varphi$ strings. We can make this more explicit by studying the fundamental group involved in the Kibble-Zurek mechanism. The potential for $\calS$ is built out of the invariants $\mathrm{tr}(\calS \calS^\dagger)$ and $\mathrm{tr}(\calS \calS^\dagger \calS \calS^\dagger)$. This potential respects a global symmetry: it is invariant under $\calS \mapsto U \calS U^\top$ with $U \in \mathrm{U}(N)$, but the element $U = -\mathds{1}_{N\times N}$ acts trivially on $\calS$. Thus, the faithfully acting global symmetry respected {\em by the potential} is $G_V = \mathrm{U}(N)/\ZZ_2$. After symmetry breaking, the subgroup of $G_V$ leaving the vacuum expectation value~\eqref{eq:Sexpsym} invariant is $H_V = \mathrm{O}(N)/\ZZ_2$. Part of these groups are gauged in our theory, but we expect that the topological defects arising when $\calS$ relaxes to the minimum of the potential are dominantly determined by $G_V$ and $H_V$, independently of the gauging. The axion string configurations formed during the phase transition are determined by $\pi_1(G_V/H_V)$. Using $\mathrm{U}(N) \cong [\mathrm{SU}(N) \times \Uone]/\ZZ_N$, we can construct noncontractible loops in $G_V$ itself as the projections of the following maps from $[0,1]$ into $\mathrm{SU}(N) \times \Uone$:
\begin{align}
  z(t) &= \left(\exp\left[-\iu t \frac{2\cpi}{N} \begin{pmatrix} 1 & & & & \\ & 1 & & & \\ & & \ddots & & \\ & & & 1 & \\ & & & & -(N-1)\end{pmatrix}\right], \exp(2\cpi \iu t/N)\right), \label{eq:zpath} \\
  w(t) &= \left(\mathds{1}, \exp(\cpi \iu t)\right). \label{eq:wpath}
\end{align}
The first path, $z(t)$, winds from the origin of $\SU(N)$ to the first element of the $\ZZ_N$ center, and also winds $1/N$ of the way around $\Uone$. The $\ZZ_N$ quotient in $U(N)$ ensures that this forms a loop in $G_V$. The action $\calS \mapsto U \calS U^\top$ implies that the corresponding winding of $\varphi$ is $4\cpi/N$. The second path, $w(t)$, winds from the identity element of $U(1)$ to the element $-1$, while remaining constant in $\SU(N)$. The $\ZZ_2$ quotient in the definition of $G_V$ makes this a loop. The action on $\calS$ corresponds to a full $2\cpi$ winding of $\varphi$. Both homotopy classes $[z]$ and $[w]$ generate $\ZZ$ subgroups of $\pi_1(G_V)$, but these are not independent, because $[z]^N$ is homotopic to $[w]^2$. We conclude that $\pi_1(G_V) \cong \ZZ$. We have $\pi_1(H_V) \cong \ZZ_2$ and $\pi_0(H_V) \cong 0$ (because the quotient by $\ZZ_2$ identifies the two disconnected components of $O(N)$). Then the homotopy exact sequence $\cdots \rightarrow \pi_1(H_V) \rightarrow \pi_1(G_V) \rightarrow \pi_1(G_V/H_V) \rightarrow \pi_0(H_V) \rightarrow \cdots$, together with the fact that the only homomorphism $\ZZ_2 \to \ZZ$ is 0, implies that $\pi_1(G_V/H_V) \cong \pi_1(G_V) \cong \ZZ$.

The physical meaning of the above analysis is that the axion strings formed by the Kibble-Zurek mechanism in this model are classified by an integer topological charge, labeled by the winding of $\varphi$, which comes in units of $2\cpi/N$. However, just as in the discussion of Sec.~\ref{sec:Zn2}, this does not mean that a string of minimal winding is an elementary object. The paths that achieve minimal winding of $\varphi$ have, from the UV viewpoint, nontrivial winding in {\em both} $\SU(N)$ and $\Uone$; they are composites of the $z(t)$ path~\eqref{eq:zpath} (analogous to the $z$ strings of Sec.~\ref{sec:Zn2}) and the $w(t)$ path~\eqref{eq:wpath} (analogous to the $\varphi$ strings of Sec.~\ref{sec:Zn2}). The string with minimal winding, then, may only exist as a collection of simpler objects, much like the depiction in Fig.~\ref{fig:2Z5_blob}. 

It seems likely that the cosmological dynamics produces primarily simple strings, $z$ strings generated by the winding~\eqref{eq:zpath} and $\varphi$ strings generated by the winding~\eqref{eq:wpath}, rather than a single string with minimal $\varphi$ winding (but much more complicated winding within $G_V$). Thus, we expect that the model has a domain wall problem, unless there is a nontrivial dynamical mechanism populating the appropriate composite strings. It would be worthwhile to explicitly construct semiclassical strings solutions of different winding, and study their relative tensions. The complete dynamics of string formation may only be definitively answered by dedicated numerical simulations.

The class of models in Ref.~\cite{Ardu:2020qmo} presents one type of embedding of $\ZZ_p$ symmetry in a continuous gauge group, and we have discussed how some nontrivial dynamical questions about the cosmic strings need to be answered before the domain wall problem can be declared solved. The models do withstand an examination of other cosmological considerations, such as existing constraints on fractionally charged relics and avoiding a Landau pole in the Standard Model couplings below the Planck scale.\footnote{There is another interesting dynamical question about the fate of $\SO(N)$ magnetic monopoles below the confinement scale, briefly discussed in Ref.~\cite{Ardu:2020qmo} but deserving further attention. For odd $N$, $\SO(N)$ has trivial center and admits $\ZZ_2$ monopoles thanks to the double cover of $\SO(N)$ by the simply connected group $\mathrm{Spin}(N)$. Such monopoles are produced as topological defects when $\SU(N)$ is higgsed to $\SO(N)$. One heuristic argument given in Ref.~\cite{Ardu:2020qmo} is that confinement is dual to higgsing, and so below the $\SO(N)$ confinement scale, the magnetic charge is higgsed and monopoles should decay. On the other hand, because there is no center symmetry, we do not expect any order parameter for confinement to exist, so it would be surprising if confinement caused monopoles to decay. The fate of the monopoles is important for the cosmology of the model, but unrelated to the concerns discussed in this paper, so we will not attempt to definitively answer the question here.} In the following sections we discuss other promising models that embed $\ZZ_p$ symmetry in more complicated symmetry structures. Some of them have greater success in solving the domain wall problem than the class of models discussed here. However, it turns out that they suffer from other cosmological considerations that cannot be simultaneously satisfied with the axion quality constraint. In other words, the axion quality problem is in tension with conventional post-inflation axion cosmology.

\section{A $\Uone$ symmetry for quality}
\label{sec:Uone}
In this section, we review a KSVZ-type model proposed by S.~M.~Barr and D.~Seckel in Ref.~\cite{Barr:1992qq}. The authors argue that this model has a high-quality $\Uone_\textsc{PQ}$ symmetry and the correct dynamics of string formation to ensure $\NDW=1$. 
In Sec.~\ref{sec:BS_review}, we review the field content and Lagrangian of the model, and discuss its solution to the quality problem. We discuss the string formation dynamics and show that the solution to the domain wall problem is much more complicated than originally claimed in \cite{Barr:1992qq}. We see that this solution reduces to a $\ZZ_p$ symmetry solution in a certain limit.  In Sec.~\ref{sec:BS_cosmo}, we show that the model also suffers from a series of cosmological problems, including fractionally charged relics and Landau poles in the SM gauge couplings below the UV scale.

\subsection{Review of the Barr-Seckel Model}
\label{sec:BS_review}

The Barr-Seckel model \cite{Barr:1992qq} introduces two complex scalars $\Phi_p$ and $\Phi_q$, which are singlets under the Standard Model gauge group. The subscripts indicate their respective charges under an additional $\Uone'$ gauge symmetry. The model also introduces $p+q$ copies of left-handed quarks (by which we mean they are in the fundamental representation of $\SU(3)_\textsc{C}$) $Q_0^{(i)}$, $q$ copies of left-handed antiquarks (by which we mean they are in the anti-fundamental representation of $\SU(3)_\textsc{C}$) $\tQ_p^{(i)}$, and $p$ copies of left-handed antiquarks $\tQ_{-q}^{(i)}$. They are singlets under $\SU(2)_\textsc{L} \times \Uone_\textsc{Y}$ and the subscripts indicate their respective $\Uone'$ charges. The multiplicities are necessitated by cancellation of $\SU(3)_\textsc{C}^3$ and $\SU(3)_\textsc{C}^2\Uone'$ anomalies. (Cancellation of the $\Uone'^3$ anomaly can be accomplished with spectator fermions that play no role in the remainder of the discussion.)

We assume that the complex scalars $\Phi_p$ and $\Phi_q$ have the appropriate potential to undergo spontaneous symmetry breaking with VEVs $v_p$ and $v_q$ respectively. Aside from the relationship between $v_p$ and $v_q$, the specific form of the scalar potential will not be relevant for our discussions. We will assume $v_p>v_q$ without loss of generality.

Given the $\Uone'$ charge assignments, the Yukawa interactions that give the quarks masses after the scalars get VEVs are therefore 
\be \label{eq:BSYukawas}
\cL_\textsc{BS}\supset (y_p)_{ij} Q_0^{(i)} \tQ_p^{(j)} \Phi_p^\dagger + (y_q)_{ij} Q_0^{(i)} \tQ_{-q}^{(j)} \Phi_q + \mathrm{h.c.}
\ee
We will denote the phase degrees of freedom of the two scalars by $\theta_p$ and $\theta_q$. Removing the $\theta_p$ degree of freedom from the Yukawa terms then requires chiral rotation of $q$ copies of $Q_0 \tQ_p$ pairs, while removing the $\theta_q$ degree of freedom requires opposite chiral rotation of $p$ copies of $Q_0 \tQ_{-q}$ pairs. The coupling of the scalar phases to the gluon is thus given by
\be
\cL_\textsc{BS} \supset \frac{q\theta_p - p \theta_q}{8 \cpi^2} \int \mathrm{tr}(G \wedge G) \equiv \frac{\theta}{8 \cpi^2} \int \mathrm{tr}(G \wedge G),
\ee
and we identify the $2\cpi$ periodic phase $\theta \equiv q\theta_p - p\theta_q$ as the axion degree of freedom that lives in $\Uone_\textsc{PQ}$, at energy scale below both $v_p$ and $v_q$. The axion decay constant is given by
\be
f_a^2  = \frac{v_p^2 v_q^2}{p^2 v_p^2 + q^2 v_q^2}.
\ee
The definition of $\theta$ makes it manifestly invariant under $\Uone'$, as it should be: the two phase rotation symmetries $\Phi_p \to \E^{\iu \theta_p} \Phi_p$ and $\Phi_q \to \E^{\iu \theta_q}\Phi_q$ have been reorganized into an anomaly-free linear combination that we have gauged and been calling $\Uone'$ and an anomalous orthogonal linear combination which we identified as $\Uone_\textsc{PQ}$. All charge assignments of the new fields are collected in Table \ref{tab:BS_originalcontent}.

We will assume that $p$ and $q$ are co-prime integers, such that the lowest order operator that breaks $\Uone_\textsc{PQ}$ symmetry is $(\Phi_p^\dagger )^{q}(\Phi_q)^p$. The correction to the axion potential induced by this operator is proportional to $v_p^q \, v_q^p / \Lambda^{p+q-4} $, where $\Lambda$ is the UV scale. The Barr-Seckel model ensures a high quality $\Uone_\textsc{PQ}$ symmetry when $p+q$ is sufficiently large. For example, for $v_p \sim v_q\sim f_a$ and the UV scale set to the Planck mass $\Lambda = M_\mathrm{Pl}$, we need $p+q \geq 10$ for $10^{10}\; \text{GeV} < f_a < 10^{12}\;\text{GeV}$ so that $\bar \theta$ will come out sufficiently small.

\begin{table}[!h]
\begin{center}
\def\arraystretch{1.2}%
\[
\begin{array}{ c|c c c c c|c } 
   & \text {spin}  & \SU(3)_\textsc{C} & \Uone_\textsc{Y} & \Uone' & \text{copies} & \Uone_\textsc{PQ}\\ 
 \hline
 Q_0  & 1/2 & \bm{3} & 0 & 0 & (p+q)& c \\ 
 \hline
 \tQ_p & 1/2 & \bm{\overline{3}}  & 0 & p & q  & a-c\\ 
 \tQ_{-q} & 1/2 & \bm{\overline{3}} & 0 & -q & p &-b-c\\ 
 \hline
 \Phi_p & 0 & \bm{1} & 0 & p & 1 & a \\
 \Phi_q & 0 & \bm{1} & 0 & q & 1 & b \\
 \hline
\end{array}
\]
\caption{\label{tab:BS_originalcontent}%
We list the beyond-SM field content of the Barr-Seckel model~\cite{Barr:1992qq}.
All the fields here are singlets of $\SU(2)_\textsc{L} \times \Uone_\textsc{Y}$.
The specific choice of $c$ does not matter, as long as the combination $Q_0 \tQ_p$ has PQ charge $a$ and $Q_0 \tQ_{-q}$ has PQ charge $-b$. Different choices of $c$ corresponds to $\Uone_\textsc{PQ}\rightarrow \Uone_\textsc{PQ} + 3c \Uone_{B}$, where $\Uone_{B}$ is the heavy baryon number. We need $p,q$ to be co-prime and $aq - bp = 1$.}
\end{center}
\end{table}

At a first glance, the setup by \cite{Barr:1992qq} seems to be exactly what we want: a high quality axion with $\NDW = 1$, since the coefficient of the $\theta$ coupling to the gluon is $1$. However, again, the problem with the construction of $\NDW = 1$ is that it's only kinematic. The fact that the minimal axion $\theta$ with minimal coupling to gluons exists does not mean that the correct string where $\theta$ winds from 0 to $2\cpi$ will form in the early universe. Indeed, from the definition of $\theta = q\theta_p - p\theta_q$, we see that $\theta$ winding from 0 to $2\cpi$ actually translates to nontrivial windings for the scalar phases $\theta_p$ and $\theta_q$. In particular, $\theta_p$ and $\theta_q$ must have higher winding numbers $a$ and $b$, such that $aq - bp = 1$. Because $p$ and $q$ are assumed to be co-prime, such integers $a$ and $b$ always exist, but the challenge is to ensure that the \textit{composite} $\Phi_p$-$\Phi_q$ string with $\theta_p$ winding number $a$ and $\theta_q$ winding number $b$ will actually form in the early universe. 

Without loss of generality, consider $v_p > v_q$. At the time of the first, $\Phi_p$ phase transition, the familiar $\Uone$ gauge strings are produced. Unless the now massive gauge boson is much heavier than the radial mode of $\Phi_p$, we expect the $\Phi_p$ strings to be dominantly winding number 1.\footnote{When the gauge boson is much heavier, we expect string-string interactions to be attractive independent of relative orientation, which may enable mergers forming higher winding strings~\cite{Laguna:1989hn, Bettencourt:1994kc, Bettencourt:1996qe}.} Therefore, we are now limited to values of $p$ and $q$ where an integer $b$ exists such that $q - bp = 1$. Later, after $\Phi_q$ undergoes its phase transition, the universe will have strings with either $\Phi_p$ or $\Phi_q$ windings or both. We will denote the strings by their $\Phi_p$ and $\Phi_q$ winding numbers, $(n,m)$, where $n$ is either $0$ or $\pm 1$, while we leave the possibility of any integer for $m$ for now. The amount of axion winding around an $(n,m)$ string is $mp-nq$, which is nonzero for generic values of $(n,m)$. This determines the number of minimal domain walls that can end on a given string, which we will refer to as the domain wall number of the string and denote $w = |mp - nq|$ (as opposed to the domain wall number $\NDW$ of the theory as a whole). In particular, $(1,m)$ strings will have both a nontrivial gauge field flux \emph{and} a nontrivial axion winding, which gives it a logarithmically divergent tension~\cite{Hill:1987bw, Klaer:2017qhr, Hiramatsu:2019tua, Hiramatsu:2020zlp, Niu:2023khv}.

In \cite{Barr:1992qq}, it is argued that at the time of the $\Phi_q$ phase transition, in the presence of a $\Phi_p$ string, only $(1,b)$ strings will form since $m = b$ minimizes the logarithmic divergence proportional to $|m - \frac{q}{p}|^2$. However, this cannot be the case, since formation of topological defects is an inherently non-local process, where defects emerge from causally disconnected patches randomly falling into different locations in the vacuum manifold. Causally disconnected regions should not be able to communicate with each other to minimize their total energy. Moreover, if there exists a process that could dynamically change the winding number to minimize energy at the time of string formation, cosmic strings would not exist at all, since the true minimal energy state is the vacuum with zero winding number.\footnote{Strings can certainly split or merge to minimize energy, but that happens strictly \textit{after} the strings have already formed from the phase transition.}

Many nontrivial questions about the $(1,b)$ string production thus remain: how many $(1,b)$ strings actually form at the time of the $\Phi_q$ phase transition? Would the abundance of $(1,b)$ string increase later due to mergers between $(1,m \neq b)$ strings with $(0,\pm 1)$ strings? Compared to pure $\Uone$ gauge strings where the kinematic constraints for mergers have been worked out in full detail \cite{Copeland:2006eh, Copeland:2006if}, mergers of strings with logarithmically divergent tensions are much less understood. Refs.~\cite{Hiramatsu:2019tua, Hiramatsu:2020zlp} have conducted simulations in the case of $p = 1$ and $q = 4$, which showed that string mergers preferentially generated $(1,4)$ strings that have domain wall number $w= 0$, instead of the desired $(1,3)$ strings with domain wall number $w=1$. However, for generic values of $p$ and $q$, there is no $(1,m)$ string with domain wall number $w=0$, and the question remains whether string mergers preferentially generate the $(1,b)$ string. Lastly, there is also the question of what is the minimal abundance of $(1,b)$ ($w = 1$) strings needed to annihilate domain walls, relative to other types of strings with $w \neq 1$. As we will see in the next section, even if all the questions about the string dynamics are answered and the domain wall problem is resolved, the model suffers from additional cosmological problems.

In the limit $v_p \gg v_q$, it is easy to see that the Barr-Seckel model solution to the axion quality problem reduces to a discrete $\ZZ_p$ symmetry. We effectively integrate out the scalar with heavier VEV, $\Phi_p$, and the gauged $\Uone'$ reduces to a $\ZZ_p$ symmetry in the vacuum. The leftover dynamical scalar $\Phi_q$ has non-minimal charge $q$ mod $p$ under the $\ZZ_p$ symmetry.
The dynamics of the string formation in this limit is exactly as discussed in Sec.~\ref{sec:Zn2}. In the simplest case of $q = 1$, the first phase transition at $v_p$ exactly produces the correct $\ZZ_p$ strings to annihilate domain walls, where the second phase transition forms strings where $p$ domain walls join. Again the model suffers from the fact that the wall-annihilating $\ZZ_p$ strings have tensions $\sim v_p$ much higher and uncorrelated with $f_a\sim v_q$, and the axion abundance from this model depends on both scales, preventing a precise prediction of the axion mass. For generic values of $q$ and $p$, as discussed in Sec.~\ref{sec:Zn2}, whether all minimal domain walls can be destroyed to prevent cosmological disaster depends on more complicated questions about string dynamics. In this limit, we would need $p$ to be large to ensure a good quality.

All the discussion above is easy to generalize to the scenario where $p$ and $q$ have greatest common divisor $d$, so that we write $p = \pt d$ and $q = \qt d$. This is very similar to the discussion in Sec.~\ref{sec:Zn}. The lowest-dimension PQ-violating operator is now $(\Phi_p^\dagger)^\qt \Phi_q^\pt$, so we need $\pt + \qt$ to be large for a high-quality $\Uone_\textsc{PQ}$ symmetry. After $\Phi_p$ and $\Phi_q$ obtain vacuum expectation values, there is a residual unbroken $\ZZ_d \subset \Uone'$ gauge symmetry, but it does not act on the axion and does not affect the domain wall problem. Effectively, the role of $p$ and $q$ in our discussion above is now played by $\pt$ and $\qt$, and the results are unchanged.

\subsection{Cosmology/quality tension}
\label{sec:BS_cosmo}
Assuming that the string dynamics already ensures that $\NDW=1$ in the Barr-Seckel model, in this subsection we analyze the problems that may still render the model invalid. These problems include fractionally charged relics in cosmology, or a more severe quality problem due to Landau poles in the SM gauge couplings below the UV scale.

New heavy quarks $Q$ introduced in KSVZ-type axion models need to be in specific representations in order to not breach the current (stringent) bounds on fractionally (electrically) charged relics \cite{DiLuzio:2017pfr,DiLuzio:2016sbl}. 
In some representations, the heavy quarks can only hadronize into fractionally charged hadrons, which cannot decay into SM particles. 
The representations with fractionally charged $Q$-hadrons are thus ruled out directly by constraints on fractionally charged relics. 
We are left to deal with only the rest of the representations in which $Q$ only form integrally charged hadrons and can decay into SM particles (equivalently, cases where the fields are in representations of $[\SU(3)_\textsc{C} \times \SU(2)_\textsc{L} \times \Uone_\textsc{Y}]/\ZZ_6$; see, e.g.,~\cite{Tong:2017oea, Davighi:2019rcd}). Furthermore, the heavy hadrons must decay sufficiently promptly to Standard Model hadrons to not violate observational constraints from BBN and CMB. Given the estimates for annihilation rates of bound states~\cite{Kang:2006yd,Jacoby:2007nw}, the authors of~\cite{DiLuzio:2017pfr,DiLuzio:2016sbl} conclude that $Q$ must have either renormalizable couplings to Standard Model quarks, or have $m_{Q}\gtrsim 800\,\text{TeV}$ for dimension-5 coupling between $Q$ and Standard Model quarks. Any higher dimensional couplings are excluded. There are only 15 representations of heavy quarks which satisfy the constraints~\cite{DiLuzio:2017pfr,DiLuzio:2016sbl}.
In particular, for $Q$ to be in the fundamental of $\SU(3)_\textsc{C}$ and to remain a singlet under $\SU(2)_\textsc{L}$, we need $\Uone_\textsc{Y}$ charge $-\frac13$ or $\frac23$, and for an anti-fundamental like $\tQ$ we need hypercharge $\frac13$ or $-\frac23$:
\begin{align} \label{eq:UoneYcharge_BS}
    Y_{Q_0} &= -\frac13 + n_1 = x, \nonumber \\
    Y_{\tQ_p} &= \frac13 - n_2 = y, \nonumber \\
    Y_{\tQ_{-q}} &= \frac13 - n_3 = z,
\end{align}
where the $n_i$'s are either 0 or 1.

One might think that once the heavy quarks have nonzero $\Uone_\textsc{Y}$ charges, the operators that allow them to decay to SM quarks will also violate $\Uone_\textsc{PQ}$ and spoil the axion solution to the strong CP problem. For example, one can write down mass terms like $m Q_0 \bar u$ and $m Q_0 \bar d$, which have nonzero PQ charge at dimension 3. However, as is argued in Ref.~\cite{Dobrescu:1996jp}, since these operators do not acquire a VEV, their modification to the axion potential is at least as suppressed as the lowest dimensional operator that involves only fields with VEVs from spontaneous $\Uone_\textsc{PQ}$ breaking. The axion solution to the strong CP problem is not spoiled in this case.  However, because of the nonzero $\Uone_\textsc{Y}$ charges the model  will run into a severe Landau pole problem, as we sketch out below.

After adding nonzero hypercharges, we now have an additional set of gauge anomaly cancellation conditions to consider. In the original model in \cite{Barr:1992qq}, the $\SU(3)_\textsc{C}^3$, $\SU(3)_\textsc{C}^2 \Uone'$, and $\SU(3)_\textsc{C} \Uone'^2$ anomalies trivially cancel. The $\Uone'^3$ anomaly can be cancelled with additional SM singlet fermions without introducing new constraints.
With the nonzero $\Uone_\textsc{Y}$ charge assignments in~\eqref{eq:UoneYcharge_BS}, we have the following gauge anomaly cancellation constraints:
\begin{align}
    \SU(3)_\textsc{C}^2 \Uone_\textsc{Y}: \quad & (p+q) x + q y + p z = 0;
    \nonumber\\
    \Uone_\textsc{Y}^3: \quad & (p+q)x^3 + q y^3 + p z^3=0;
    \nonumber\\
    \Uone_\textsc{Y}^2 \Uone': \quad & q y^2 p + p z^2 (-q) = 0;
    \nonumber\\
    \Uone_\textsc{Y} \Uone'^2: \quad & q y p^2 + p z q^2 = 0.
\end{align}
The fields $\Phi_p$ and $\Phi_q$ must have zero hypercharge, because they obtain large VEVs. Then the structure of the Yukawa couplings in~\eqref{eq:BSYukawas} implies that we must take $x = -y = -z$. All anomaly constraints except the last one (and including the $\Uone_\textsc{Y}$ gravitational anomaly) are immediately satisfied after this assumption. The last constraint, which becomes $(qp^2 + pq^2)y = 0$, cannot be satisfied with the field contents described above. Instead, we have to enlarge the field content of the theory. If we simply double the fermion content of Table~\ref{tab:BS_originalcontent}, we would have to give opposite hypercharges to the two copies. This is inconsistent with the quantization rule $y = 1/3 + \ZZ$; it would also double the $\SU(3)_\textsc{C}^2 \Uone_\textsc{PQ}$ anomaly, spoiling this model as a solution to the domain wall problem.

The minimal solution is to have {\em three} times the original fermion content. To declutter our notation, we suppress the original copy indices $i, j$ from~\eqref{eq:BSYukawas}, and keep only an index $a \in \{1, 2, 3\}$ to label the new copies or generations of the fields. We take the heavy generations to have hypercharges $x^a, y^a, z^a$. All anomaly constraints are then satisfied with $x^a = -y^a = -z^a$ and $\sum_a y^a = 0$. In order to maintain domain wall number $\NDW = 1$, we take the third generation to have the opposite-sign $\Uone'$ charge (and, correspondingly, an opposite-sign $\Uone_\textsc{PQ}$ charge) from the first two, so that it contributes an equal and opposite amount to the $\SU(3)_\textsc{C}^2\Uone_\textsc{PQ}$ anomaly. The Yukawa couplings then take the form
\begin{align}
\label{eq:BS_add_lagrangian}
    \cL \supset \sum_{a=1,2} \Bigl( y_p^a Q_0^a \tQ_p^a \Phi_p^\dagger + y_q^a Q_0^a \tQ_{-q}^a \Phi_q \Bigr)
    + y_p^3 Q_0^3 \tQ_p^3 \Phi_p + y_q^3 Q_0^3 \tQ_{-q}^3 \Phi_q^\dagger
    + \mathrm{h.c.}
\end{align}
This argument can be understood by thinking about the chiral rotation of the $Q_0 \tQ_p$ combination in the interaction term $Q_0 \tQ_p \Phi_p^\dagger$. In order to cancel the $\SU(3)_\textsc{C}^2\Uone_\textsc{PQ}$ anomaly contribution from rephasing $\Phi_p$ between two generations, we need one generation to couple to $\Phi_p^\dagger$ and the other to couple to $\Phi_p$. This coupling is determined by the $\Uone'$ charges of the $Q_0$ and $\tQ_p$ fields. The same logic follows for the coupling to $\Phi_q$. Since the axion is a linear combination of the rephasing of $\Phi_p$ and $\Phi_q$, the above treatment cancels the $\SU(3)_\textsc{C}^2\Uone_\textsc{PQ}$ anomaly between two generations, and the remaining generation contributes $\NDW = 1$.

We choose the smallest possible hypercharge assignments to make the Landau pole constraint as mild as possible, choosing $n_i = 0$ in~\eqref{eq:UoneYcharge_BS} for the first two generations and $n_i = 1$ for the third generation to achieve $\sum_a y^a = 0$. 
The additional matter field content with the fields' representations is displayed in Table~\ref{tab:content_BS} for the readers' convenience.

\begin{table}[!h]
\begin{center}
\def\arraystretch{1.5}%
\[
\begin{array}{ c| c c c c } 
    & \SU(3)_\textsc{C} & \Uone_\textsc{Y} & \Uone' &\text{copies} 
   \\ 
 \hline
 Q_0^i & \bm{3} & x^i= (-\frac13, -\frac13, \frac23) & (0,0,0) &(p+q)\times 3 \\ 
 \hline
 \tQ^i_p & \bm{\overline 3} & y^i= (\frac13, \frac13, -\frac23)& (p,p,-p) & q\times 3  \\ 
 \tQ^i_{-q} & \bm{ \overline 3} & z^i= (\frac13, \frac13, -\frac23) & (-q,-q,q)  & p\times 3\\ 
\hline
\end{array}
\]
\caption{\label{tab:content_BS}%
We list the modified fermion contents of our adaptation of the model in Ref.~\cite{Barr:1992qq}.
The fields are still singlets under $\SU(2)_\textsc{L}$, but they now acquire nontrivial $\Uone_\textsc{Y}$ charge.
The three generations of fields are required to cancel gauge anomalies as well as to maintain $\NDW=1$. The $\Uone_\textsc{PQ}$ charge closely resembles that in Table \ref{tab:BS_originalcontent}. The quark combination $Q_0^3 \tQ_p^3$ in the third generation has opposite $\Uone_\textsc{PQ}$ charge from the first two, and the interaction with the scalars is correspondingly modified. 
All the fields in this table are left-handed Weyl fermions.
}
\end{center}
\end{table}

Our next task is to assess the quality problem in light of the UV cutoff of our effective field theory. Before getting to the detailed calculations, let us explain the logic behind our argument. We expect that the PQ symmetry, like all global symmetries, is badly violated at or below the fundamental UV cutoff energy $\Lambda_\textsc{QG} \lesssim M_\mathrm{Pl}/\sqrt{N}$ (where $N$ is the number of weakly-coupled degrees of freedom below the cutoff~\cite{Veneziano:2001ah}), at which gravity becomes strongly coupled and local quantum field theory breaks down. To assess the quality problem, we assume that the EFT below $\Lambda_\textsc{QG}$ includes all PQ-violating operators suppressed by $\Lambda_\textsc{QG}$. (In Sec.~\ref{sec:discussion}, we will briefly comment on the possibility that some mechanism causes the coefficients of these operators to be much more suppressed.) The large matter content of our theory will lead to a Landau pole for Standard Model gauge couplings, at a UV cutoff scale $\Lambda_\textsc{SM}$. In general, $\Lambda_\textsc{SM}$ and $\Lambda_\textsc{QG}$ are different scales. There are two cases to consider. First, suppose that $\Lambda_\textsc{SM} < \Lambda_\textsc{QG}$. Then, in the energy range $\Lambda_\textsc{SM} \lesssim E \lesssim \Lambda_\textsc{QG}$, physics is still described by local quantum field theory, but  this theory is not the weakly coupled gauge theory that we started with. It may be a different weakly coupled gauge theory (as in Seiberg duality), or it may not have any weakly coupled description. It is possible that PQ-violating operators in the EFT below $\Lambda_\textsc{SM}$ are not suppressed by $\Lambda_\textsc{SM}$, because PQ breaking may only arise at the higher scale $\Lambda_\textsc{QG}$. However, we do not {\em know} the description of the theory in this window, and we cannot enumerate $\Lambda_\textsc{QG}$-suppressed PQ-breaking operators above $\Lambda_\textsc{SM}$, match them onto operators below $\Lambda_\textsc{SM}$, and quantify the amount of PQ symmetry violation in the low-energy theory. Thus, from our viewpoint, in this scenario we can never claim to have solved the axion quality problem. We will thus, instead, assume that $\Lambda_\textsc{QG} < \Lambda_\textsc{SM}$. In that case, we can assume that the EFT we are working with is valid all the way up to a scale $\Lambda$ that suppresses PQ-violating operators, and we can assess whether this is sufficient to solve the problem. Let us emphasize that our logic is that we are testing whether we have a theory in which we can definitely claim that the quality problem has been solved. We are {\em not} attempting to prove a no-go theorem, which is a more difficult task that would require us to address the other ordering $\Lambda_\textsc{SM} < \Lambda_\textsc{QG}$. It would be interesting to attempt to find an explicit example of this type in which we can extend our low-energy theory, through duality, to include the high-energy regime, and test whether the quality problem can be solved. To find a controlled example, we would likely have to assume approximate supersymmetry at the scale $\Lambda_\textsc{SM}$. Constructing such theories is beyond the scope of this paper.

To assess the tension between axion quality and Landau pole constraints in the scenario with $\Lambda_\textsc{QG} < \Lambda_\textsc{SM}$, we use the one-loop $\beta$-functions of $\SU(3)_\textsc{C}$ and $\Uone_\textsc{Y}$ to determine the location of the Landau poles.
Consider the $\SU(3)_\textsc{C}$ running coupling at one-loop:
\begin{align}
    \frac{1}{\alpha_s(\mu)} = \frac{1}{\alpha_s(\mu_0)} - \frac{b_0}{2\cpi} \ln{\frac{\mu}{\mu_0}}
\,,\end{align}
where $b_0 = -11 + \frac{2}{3}n_f$.
In the following we assume that the Yukawa couplings of the quarks to the scalars are 1. Decreasing the Yukawa couplings and thus the heavy quark masses will make the Landau pole problem worse since the heavy quarks start affecting the running of the coupling at a lower scale. We have checked that increasing the Yukawa coupling to the maximum possible value of $4\cpi$ will not change the conclusion of our analysis either. We run the coupling from mass of the top quark $m_t$ to the smaller VEV of the two scalars, $v_q$, with SM fields only. 
Above $v_q$ and below $v_p$ we have $3\times p$ additional \emph{Dirac} pairs of $(\tQ_{-q}, Q_0)$ fields,
and above $v_p$ we have another $3\times q$ additional $(\tQ_{p},Q_0)$ fields. 
Requiring the $\SU(3)_\textsc{C}$ Landau pole to be below the new physics scale $\Lambda$, we need
\begin{align} \label{eq:BS_SU3}
    -7 \ln \frac{m_t}{\Lambda} + 2 \Bigl(q \ln \frac{v_p}{\Lambda} + p \ln \frac{v_q}{\Lambda}\Bigr) + 
    \frac{2 \cpi}{\alpha_s(m_t)}
    > 0
\,.\end{align}
Similarly, we can derive the constraints for the $\Uone_\textsc{Y}$ Landau pole:
\begin{align}\label{eq:BS_U1Y}
    \frac{41}{6} \ln \frac{m_t}{\Lambda} + 4 \sum_{i} Q_{i,Y}^2 \Bigl(q \ln \frac{v_p}{\Lambda} + p \ln \frac{v_q}{\Lambda}\Bigr) + \frac{2 \cpi}{\alpha_\textsc{Y}(m_t)}> 0 
\,,\end{align}
where $\sum_{i} Q_{i,Y}^2 = \sum_{i} x_i^2 = \sum_{i} y_i^2 = \sum_{i} z_i^2 = \frac23$ is the sum over squares of $\Uone_\textsc{Y}$ charges for each quark field for all 3 generations.
Finally, the quality problem constraint is that the correction to axion potential from operators of the form
$g (\Phi_p^\dagger)^q (\Phi_q)^p / \Lambda^{p+q-4}$ is smaller than $10^{-10} \Lambda_\textsc{IR}^4$, where $\Lambda_\textsc{IR}^4 = \frac{m_u m_d}{(m_u+m_d)^2} m_\pi^2 f_\pi^2$ is the size of the QCD-generated contribution to the axion potential.
We then require when $\Phi_p$ and $\Phi_q$ get VEVs,
\begin{align} \label{eq:BS_quality}
    \ln g + \Bigl(q \ln \frac{v_p}{\Lambda} + p \ln \frac{v_q}{\Lambda}\Bigr) - 4 \ln \frac{\Lambda_\textsc{IR}}{\Lambda}  
    < \ln 10^{-10}
\,.\end{align}

\begin{figure}[!ht]
\centering
\includegraphics[width=0.5\textwidth]{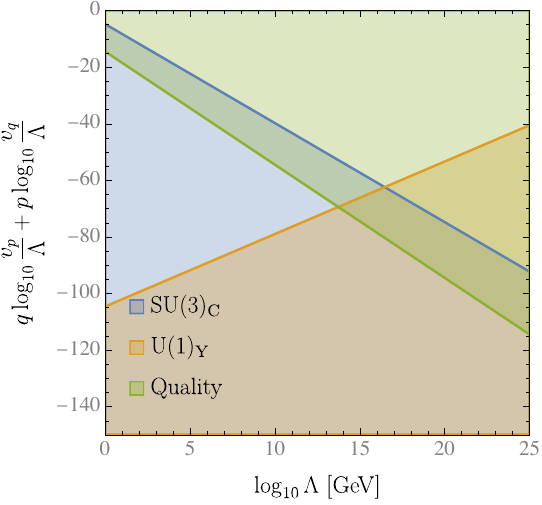}
\caption{We show that the three constraints from Standard Model Landau poles and axion quality exclude the parameter space of the model in Sec.~\ref{sec:BS_cosmo} completely. The shaded region for each color indicates the excluded region by imposing the given constraint, i.e. the blue (orange) region indicates where the $\SU(3)_\textsc{C}$ ($\Uone_\textsc{Y}$) Landau pole is below the UV scale $\Lambda$, and the green region is where the correction to the axion potential is larger than $10^{-10}$. In the plot the Yukawa couplings and the coupling strength of the lowest-dimensional PQ breaking operator $g$ are set to 1, but varying the couplings does not change the conclusion unless $g$ is taken extremely small. We do not assume the value any other parameters in the model, as the dependence can be fully captured by the combination on the $y$-axis, which is $q \log_{10} \frac{v_p}{\Lambda} + p \log_{10} \frac{v_q}{\Lambda}$ for the model in Sec.~\ref{sec:BS_cosmo}. We later discuss a different model in Sec.~\ref{sec:SUnSUn}, and find similar conclusions. The $y$-axis dependence in that case becomes $n  \log_{10} \frac{m_Q}{\Lambda}$. See the text for more detail.}
\label{fig:exlusion}
\end{figure}

We observe that the combination $q \ln \frac{v_p}{\Lambda} + p \ln \frac{v_q}{\Lambda}$ is the defining parameter of this family of models.
Combining the three constraints into one exclusion plot shown in Fig.~\ref{fig:exlusion}, where we take $g = 1$, we see that the parameter space is completely ruled out. 

One might ask how small $g$ would need to be to avoid this conclusion. Combining~\eqref{eq:BS_quality} and~\eqref{eq:BS_SU3}, we derive
\begin{equation}
    g < 10^{-10} \left(\frac{\Lambda_\textsc{IR}}{m_t}\right)^{7/2} \sqrt{\frac{\Lambda_\textsc{IR}}{\Lambda}} \E^{\pi/\alpha_s(m_t)} \approx 10^{-17} \sqrt{\frac{10^{12}\,\mathrm{GeV}}{\Lambda}}.
\end{equation}
This shows that, for any reasonable cutoff on a QCD axion model, solving the quality problem while avoiding a Landau pole for QCD requires that the coefficient of the lowest-dimension PQ-violating operator be tuned much smaller than $\thetabar$ itself. Thus, the Strong CP problem is simply not solved, but (at best) transformed into a similar problem of why $g$ is so small.

\section{An $\SU(n) \times \SU(n)$ symmetry for quality}
\label{sec:SUnSUn}

In this section we will discuss a model of a high quality axion protected by a nonabelian gauge symmetry, following~\cite{DiLuzio:2017tjx}. Many of the considerations are closely parallel to those in the previous section, so our discussion will be relatively brief. 

\subsection{Scalar potential and axion strings}

The theory has a gauge symmetry $G = \SUNL \times \SUNR$, with a field $Y$ in the bifundamental $(\Yfund, \overline{\Yfund})$ representation. The renormalizable potential $V(Y)$ contains a tachyonic $\mathrm{tr}(Y Y^\dagger)$ term and quartic terms which are a linear combination of $\mathrm{tr}(Y Y^\dagger Y Y^\dagger)$ and $\mathrm{tr}(Y Y^\dagger)^2$, which lead to a vacuum expectation value $\langle Y \rangle \propto \mathds{1}$ breaking $\SUNL \times \SUNR$ to the diagonal $\SUNV$. This symmetry breaking pattern is familiar from the flavor symmetry of QCD, and the structure of the potential for $Y$ mimics that of the  linear sigma model. The renormalizable potential also has an accidental $\Uone$ global symmetry rephasing $Y$, which is of high quality; the lowest-dimension operator breaking it is $\det Y$. This is the Peccei-Quinn symmetry that is spontaneously broken to provide an axion,
\begin{equation}
    \theta(x) = \arg(\det Y(x)).
\end{equation}

To understand the formation of axion strings, consider the limit in which the gauge couplings are turned off and we focus on the dynamics induced by the scalar potential. The renormalizable potential has a global symmetry
\begin{equation} \label{eq:GV}
    G_V = [\SUNL \times \SUNR \times \Uone]/(\ZZ_n \times \ZZ_n).
\end{equation}
Here $\SUNL: Y \mapsto L Y$, $\SUNR: Y \mapsto Y R^\dagger$, and $\Uone: Y \mapsto \E^{\iu \alpha} Y$. The $\ZZ_n$ group generated by $L = R = \exp(2\cpi \iu/n)\mathds{1}$ and $\alpha = 0$ acts trivially on $Y$, as does the $\ZZ_n$ group generated by $L = \exp(2\cpi \iu/n)\mathds{1}, R = \mathds{1}, \alpha = -2\cpi /n$, so we have taken the quotient by the product of these two $\ZZ_n$ groups to obtain the group $G_V$ acting faithfully on $Y$.

The vacuum expectation value $\langle Y \rangle = v \mathds{1}$ spontaneously breaks the symmetry to 
\begin{equation} \label{eq:HV}
H_V = \mathrm{PSU}(n)_\textsc{V} \equiv \SUNV / \ZZ_n.
\end{equation}
That is, pairs $L = R = V \in \SUN$ acting as $Y \mapsto V Y V^\dagger$ leave $\langle Y \rangle$ invariant, but the cases where $V$ is in the center of $\SUN$ acted trivially on {\em any} $Y$ and were quotiented out in $G_V$, so we must take the $\ZZ_n$ quotient here as well.

The phase transition in which $Y$ acquires a vacuum expectation value will lead, by the Kibble-Zurek mechanism, to the production of strings characterized by the fundamental group $\pi_1(G_V / H_V)$. We claim that $\pi_1(G_V / H_V) \cong \ZZ$. To see this, we can realize $G_V$ as a quotient of the simply connected space $\SUNL \times \SUNR \times \RR$ by the freely-acting group $\ZZ_n \times \ZZ$, where $\ZZ_n$ is generated by $(L, R, \alpha) \mapsto (\E^{2\cpi \iu/n} L, \E^{2\cpi \iu/n} R, \alpha)$ and $\ZZ$ is generated by $(L, R, \alpha) \mapsto (\E^{2\cpi \iu/n} L, R, \alpha - 2\cpi/n)$. Hence $\pi_1(G_V) \cong \ZZ_n \times \ZZ$. We also have $\pi_1(H_V) \cong \ZZ_n$. The inclusion $H_V \hookrightarrow G_V$ induces a map identifying $\pi_1(H_V)$ with the $\ZZ_n$ factor in $\pi_1(G_V)$. We use the homotopy exact sequence $\ldots \to \pi_1(H_V) \to \pi_1(G_V) \to \pi_1(G_V/H_V) \to \pi_0(H_V) \cong 0$ to conclude that $\pi_1(G_V/H_V) \cong \ZZ$. This fundamental group is generated by the image of the path $[0,1] \to \SUNL \times \SUNR \times \Uone$ given by
\begin{equation}
t \mapsto \left(\exp\left[\iu t \frac{2\cpi}{n} \begin{pmatrix} 1 & & & & \\ & 1 & & & \\ & & \ddots & & \\ & & & 1 & \\ & & & & -(n-1)\end{pmatrix}\right], \mathds{1}, \exp(-2\cpi \iu t / n)\right).
\end{equation}
In the covering space, this path winds from the origin to a central element in $\SUNL$ and winds $1/n$ of the way around $\Uone$; this endpoint is identified with the origin by the second $\ZZ_n$ factor in~\eqref{eq:GV}. The Kibble-Zurek mechanism leads to the production of cosmic strings associated with this path, which correspond to strings of minimal winding for the axion $\theta$. We do not expect that gauging $\SUNL \times \SUNR$ and adding Yukawa interactions will alter this conclusion. Unlike for the models of Sec.~\ref{sec:Zn3} and Sec.~\ref{sec:Uone}, then, there is no ambiguity about string formation dynamics in this model: the domain wall problem can be solved, if the model is free of other cosmological difficulties.

\subsection{Standard Model couplings and Landau pole constraint}

The model of~\cite{DiLuzio:2017tjx} adds quark fields charged under $\SUNL \times \SU(3)_\textsc{C}$ and $\SUNR \times \SU(3)_\textsc{C}$, which give rise through triangle diagrams to the couplings of the axion to gluons. These fields were chosen to be electroweak singlets, and so~\cite{DiLuzio:2017tjx} assumes a pre-inflation scenario to avoid the cosmological disaster of fractionally charged particles. However, as in Sec.~\ref{sec:BS_cosmo}, we can consider a model in the post-inflation scenario with additional fields carrying hypercharge to eliminate this problem. 

\begin{table}[!h]
\begin{center}
\def\arraystretch{1.2}%
\[
\begin{array}{ c|c c c c c c | c } 

   & \text{spin} & \SUNL & \SUNR & \SU(3)_\textsc{C} & \Uone_\textsc{Y} & \text{copies} & \Uone_\textsc{PQ} \\ 
\hline
Y & 0 & \Yfund & \overline{\Yfund} & \bm{1} & 0 & 1 & 1/n \\
 \hline
 D^{(i)} & 1/2 & \Yfund & \bm{1} & \bm{3} & -1/3 & 2 & 1/n \\
 \widetilde{D}^{(i)} & 1/2 & \bm{1} & \overline{\Yfund} & \bm{\overline{3}} & +1/3 & 2 & 0 \\
 \hline
 U & 1/2 & \overline{\Yfund} & \bm{1} & \bm{3} & +2/3 & 1 & -1/n \\
 \widetilde{U} & 1/2 & \bm{1} & \Yfund & \bm{\overline{3}} & -2/3 & 1 & 0 \\ 
 \hline
 \Psi^{(j)} & 1/2 & \overline{\Yfund} & \bm{1} & \bm{1} & 0 & 3 & -1/n \\
 \widetilde{\Psi}^{(j)} & 1/2 & \bm{1} & \Yfund & \bm{1} & 0 & 3 & 0\\
 \hline
\end{array}
\]
\caption{\label{tab:SUnSunmodel}%
Matter content of a model with an $\SUN \times \SUN$ gauge symmetry to address the quality problem. This closely follows~\cite{DiLuzio:2017tjx}, but we have added nontrivial hypercharge assignments to avoid fractionally charged particles. This then requires a more elaborate field content, chosen to cancel gauge anomalies and to ensure domain wall number 1. All spin-$1/2$ fields in table are left-handed Weyl fermions.}
\end{center}
\end{table}

Specifically, we consider the matter content shown in Table~\ref{tab:SUnSunmodel}, together with the Yukawa couplings
\begin{equation}
    (y_D)_{ik} D^{(i)}Y^\dagger \widetilde{D}^{(k)} + y_U U Y \widetilde{U} + (y_\Psi)_{jl} \Psi^{(j)} Y \widetilde{\Psi}^{(l)} + \mathrm{h.c.}
\end{equation}
The symmetry $\Uone_\textsc{PQ}$ is an approximate accidental global symmetry; we have normalized the charges to be fractional to signal that the $2\cpi$-periodic axion field $\theta$ is the phase of $\det Y$, rather than $Y$ itself. This gives rise to the correctly normalized domain wall number from the $\SU(3)_\textsc{C}^2\text{-}\Uone_\textsc{PQ}$ anomaly coefficient: $2 \times n \times 1/n + n \times (-1/n) = 1$. Importantly, the $\SUN_\textsc{L,R}^2\text{-}\Uone_\textsc{PQ}$ mixed anomalies cancel, ensuring that the new nonabelian dynamics does not spoil the solution to the Strong CP problem. Aside from $\Uone_\text{PQ}$, all other symmetries in the table are gauged, and the charges are chosen so that anomalies cancel. For example, the mixed $\SUNL^2\text{-}\Uone_\textsc{Y}$ anomaly cancels because we have two copies of the $D$ fields with hypercharge $-1/3$, but only one copy of the $U$ field with hypercharge $+2/3$. The $\SUNL^3$ anomaly cancels because we have 2 copies of a color triplet in the $\Yfund$ representation and one color triplet and three copies of a singlet in the $\overline{\Yfund}$ representation, so altogether there are six fundamentals and six antifundamentals. Similarly, one can check that all of the other gauge anomaly cancellation conditions hold.

Much as in Sec.~\ref{sec:BS_cosmo}, the large number of added fields poses a Landau pole problem. We assume that all the additional fermion fields get mass $m_Q \sim v/\sqrt{2}$. This is valid if all Yukawa couplings are of the same order. The masses are already naturally degenerate within each $\SU(3)_\textsc{C}$ and $\SU(n)_\textsc{L+R}$ multiplet. 
Following a similar analysis as before, in which the fermions alter the beta functions above the scale $m_Q$,
we arrive at the Landau pole constraints for $\SU(3)_\textsc{C}$ and $\Uone_\textsc{Y}$:
\begin{align}
    -7 \ln \frac{m_t}{\Lambda} + 2 n \ln \frac{m_Q}{\Lambda} + 
    \frac{2 \cpi}{\alpha_s(m_t)}
    > 0\,,  \label{eq:dL_SU3}
    \\
    \frac{41}{6} \ln \frac{m_t}{\Lambda} + 4n \sum Q_{i,Y}^2 \ln \frac{m_Q}{\Lambda} + \frac{2 \cpi}{\alpha_\textsc{Y}(m_t)}> 0 \,. \label{eq:dL_U1Y}
\end{align}
Here $\sum Q_{i,Y}^2 = \frac{2}{3}$ sums over $D^{(i)}, U, \Psi^{(j)}$, or equivalently $\widetilde{D}^{(i)},  \widetilde{U},  \widetilde{\Psi}^{(j)}$.

The quality problem constraint comes from requiring the correction to the axion potential 
\cite{DiLuzio:2017tjx}
\begin{align}
    \Delta V = \Big(\frac{v}{\sqrt 2}\Big)^n \frac{\kappa}{\Lambda^{n-4}}
\end{align}
to satisfy $\Delta V/\Lambda_\textsc{IR}^4 < 10^{-10}$. Here $\kappa$ is the coupling strength of the lowest-dimensional PQ breaking operator $\det Y$, and we take $\kappa$ to be $\mathcal{O} (1)$.
This translates to
\begin{align}\label{eq:dL_quality}
    \ln \kappa + n \ln \frac{m_Q}{\Lambda} - 4 \ln \frac{\Lambda_\textsc{IR}}{\Lambda} < \ln 10^{-10} \,,
\end{align}
where we have used $m_Q \sim v/\sqrt{2}$. 

The exclusion plot for this model looks the same as Fig.~\ref{fig:exlusion} from the model in Sec.~\ref{sec:BS_cosmo}, except the defining parameter on the $y$-axis is now $n \ln \frac{m_Q}{\Lambda}$. This is clear to see by comparing the constraint equations themselves.
We conclude that this class of models with nonabelian gauge symmetry also presents tensions between the quality problem constraint and standard post-inflation cosmology. We reiterate that the domain wall problem is indeed solved here, as the string formation dynamics is unambiguous, unlike the models considered in Sec.~\ref{sec:Zn} and Sec.~\ref{sec:Uone}, where it remains unclear without detailed simulation whether the correct strings can form to annihilate domain walls. Nonetheless, the model fails either because of fractionally charged relics or because it does not solve the quality problem (and hence, does not solve the Strong CP problem).

\section{Comments on composite axion models}
\label{sec:composite}

The models that we have discussed above have elementary scalar fields, which introduce a fine tuning problem of their own. This scalar hierarchy problem could be solved by supersymmetry or compositeness. It has long been appreciated that a composite axion model could naturally explain the axion's origin as a pseudo-Nambu-Goldstone boson, much like the pions of QCD~\cite{Kim:1984pt}. The minimal such models are free of a scalar hierarchy problem, but do not solve the quality problem because they admit low-dimension PQ-violating operators. Over the years, a number of examples have been constructed in which additional gauge symmetries forbid such operators and solve the quality problem; see, e.g.,~\cite{Randall:1992ut, Dobrescu:1996jp, Redi:2016esr, Gavela:2018paw, Contino:2021ayn, Nakagawa:2023shi}. These ensure that the lowest-dimensional PQ-violating operator has a large dimension by realizing it as a baryonic operator of some $\SU(m)$ gauge theory (e.g.,~\cite{Randall:1992ut}) or as a product across many links in a moose diagram (e.g.,~\cite{Redi:2016esr}). 

All of these models feature a confining $\SU(n)$ gauge theory, and all of them predict a domain wall number that is a multiple of $n$. Thus, none of them is a good candidate for a post-inflation axion realizing the minimal solution to the domain wall problem.

The reason that the domain wall number is a multiple of $n$ in these models is that the heavy fields carrying PQ charge are fundamentals of both $\SU(n)$ and $\SU(3)_\textsc{C}$. Integrating out fields carrying PQ charge $p_i$, $\SU(n)$ representation $r_i$, and $\SU(3)_\textsc{C}$ representation $s_i$ gives
\begin{equation} \label{eq:NDWformula}
k_G = \sum_i p_i \dim(r_i) 2I(s_i),
\end{equation}
with $I(s_i)$ the Dynkin index normalized to $1/2$ for the fundamental representation. For the moment, we assume that $p_i \in \ZZ$ (an assumption that we will revisit below). We have $2I(s_i) \in \ZZ$, and for the fundamental representation $\Yfund$ of $\SU(n)$, $\dim(\Yfund) = n$. From this expression, we see that in order to modify a composite axion model to have $|k_G| = 1$, we must necessarily have fields contributing to $k_G$ that transform in different representations of $\SU(n)$.

Let us consider a composite model with the following assumptions:
\begin{itemize}
\item It is based on strong dynamics arising from an asymptotically free $\SU(n)$ gauge theory.
\item All of the PQ-charged fields contributing to $k_G$ transform nontrivially under $\SU(n)$.
\item The PQ symmetry has no mixed anomaly with $\SU(n)$; otherwise, the would-be axion would, like the $\eta'$ in QCD, acquire a large mass.
\end{itemize}
We will see that, even before considering dynamics or how to arrange for a high-quality accidental PQ symmetry, there are some quite general difficulties with achieving domain wall number 1 from this starting point.

Asymptotic freedom of $\SU(n)$ is a strong restriction on the matter content. At one loop, it requires that
\begin{equation}
\frac{11}{3} n - \frac{1}{3} \sum_i 2I(r_i) \dim(s_i) > 0,
\end{equation}
where the sum is over left-handed Weyl fermions $\psi_i$ transforming in the $r_i$ representation of $\SU(n)$ and the $s_i$ representation of $\SU(3)_\textsc{C}$. For a nontrivial representation of $\SU(3)_\textsc{C}$ we have $\dim(s_i) \geq 3$. Hence, only representations satisfying
\begin{equation}
2I(r_i) < \frac{11}{3} n
\end{equation} 
can preserve asymptotic freedom. This eliminates large representations from consideration. For example, the symmetric three-index tensor has $2I(\Ythrees) = (n+2)(n+3)/2$, which exceeds $11n/3$ for all $\SU(n)$. The fundamental, two-index tensors, and adjoint always obey the inequality. Other tensors do in special cases, e.g., the fully antisymmetric three-index tensor ($2I(\Ythreea) = (n-3)(n-2)/2$) when $n < 12$, the mixed three-index tensor ($2 I(\Yadjoint) = n^2 - 3$) when $n = 4$, and even the four-index fully antisymmetric tensor when $n = 8$. 

Simply having multiple representations of $\SU(n)$ available is no guarantee that we can find a viable model with domain wall number 1. For example, any model with only fundamentals and symmetric or antisymmetric two-index tensor representations will have non-minimal domain wall number. This is because, for each of these representations $r$, we have $n \mid \dim(r)$ when $n$ is odd, and $\frac{n}{2} \mid \dim(r)$ when $n$ is even. In fact, we can make a stronger statement. A representation $r$ of $\SU(n)$ has an associated $n$-ality $z_n(r)$, which is the representation's charge under the $\ZZ_n$ center of $\SU(n)$, or equivalently the number of boxes in the Young tableau modulo $n$. Any representation has the property that
\begin{equation} \label{eq:zndimprop}
z_n(r) \dim(r) \cong 0 \pmod n.
\end{equation}
This follows from the fact that for any $\SU(n)$ representation $\rho$, $\det\rho(g) = 1$ for any $g \in \SU(n)$ (and in particular, for $g$ in the center); this is because $\det \circ \rho$ gives a one-dimensional representation of $\SU(n)$ and all such representations are trivial.
Using~\eqref{eq:zndimprop}, we can  exclude many possible models in which all of the $\dim(r)$ terms in~\eqref{eq:NDWformula} have certain factors of $n$ in common. For example, if $n$ is a prime power $p^k$, then the dimension of {\em every} representation of nonzero $n$-ality is divisible by $p$, and achieving domain wall number 1 requires the use of a representation of zero $n$-ality (like the adjoint). 

For an example where $n$ is not a prime power, consider the case of $\SU(6)$. We have $\dim(\Yfund) = 6$, $\dim(\Yasymm) = 15$, and $\dim(\Ythreea) = 20$. Any model that exploits only two of these representations has non-minimal domain wall number (divisible by 3 using $\Yfund$ and $\Yasymm$, by 2 using $\Yfund$ and $\Ythreea$, or by 5 using $\Yasymm$ and $\Ythreea$). A model that exploits all three representations (or their conjugates) has a chance. However, such a model has a very large amount of matter! If we add fields in the $(\Yfund, \Yfund)$, $(\Yasymm, \Yfund)$, and $(\Ythreea, \Yfund)$ of $\SU(6) \times \SU(3)_\textsc{C}$ (possibly with conjugates on some of these labels), we have added $3(6 + 15 + 20) = 123$ new fields that must carry hypercharge (to avoid fractional charges, as usual). If we assume that all of these fields have the minimal hypercharge $\pm 1/3$ compatible with their $\SU(3)_\textsc{C}$ representation, then they are sufficient to drive $\Uone_\textsc{Y}$ to a Landau pole below the Planck scale unless $f_a \gtrsim 4 \times 10^{15}\,\mathrm{GeV}$, which would lead to an overabundance of dark matter in the post-inflation scenario. In fact, this field content is insufficient: we must cancel the $\Uone_\textsc{Y}\text{-}\SU(6)^2$ anomaly, among others, so the situation can only become worse. And so far we have only discussed field content; we must also ask whether the model admits an approximate Peccei-Quinn symmetry of high quality, what the PQ charge assignments are, and whether the model has the appropriate dynamics to spontaneously break PQ without unwanted cosmological relics! It seems likely that, even if we lower the cutoff below the Planck scale, no model along these lines could be viable.

Next, let us consider the case of the adjoint representation. We have $2I(\mathrm{Adj}) = 2n$, so we can add a field $\lambda$ that is an $(\mathrm{Adj},\Yfund)$ of $\SU(n) \times \SU(3)_\textsc{C}$ without spoiling the asymptotic freedom of $\SU(n)$, but we cannot add two of them. We could also add $(\Yfund, \Yfund)$ or $(\Yfund, \overline{\Yfund})$ representations (at least one, possibly more, depending on Peccei-Quinn charge assignments) to achieve domain wall number 1. Along the lines of the previous paragraph, we have added at least $3(n^2 + n - 1)$ new fields that must carry hypercharge at least $\pm 1/3$, so $n$ cannot be too large. For example, demanding a Landau pole below the Planck scale for $f_a = 10^{11}\,\mathrm{GeV}$, we must have $n < 7$. One could further consider anomaly cancelation and other constraints on the model,  but there is another problem to consider that is specific to this case. The adjoint is a real representation, and there is no reason to expect that $\langle \mathrm{tr}(\lambda \lambda) \rangle$ should be zero. This expectation value would spontaneously break $\SU(3)_\textsc{C}$ and $\Uone_\textsc{Y}$ at the scale $f_a$, in obvious contradiction to the world around us.

One of the assumptions that we made above, integer PQ charges $p_i \in \ZZ$, can be modified in cases where a discrete subgroup of $\Uone_\textsc{PQ}$ is gauged. As we emphasized at the end of Sec.~\ref{sec:Zn2}, in such cases, the formula~\eqref{eq:NDWformula} that we have used to compute $k_G$ may overcount the true number of gauge-inequivalent vacua. Instead, we should divide by the number of vacua connected by discrete gauge transformations acting on the axion. In other words, the Lazarides-Shafi mechanism~\cite{Lazarides:1982tw} gives a potential loophole in the above argument, just as it did in our examples in previous sections. A gauged discrete subgroup of $\Uone_\textsc{PQ}$ cannot be embedded in the confining $\SU(n)$ group itself, because we have assumed that the axion is a meson of the confining sector (and hence that it is invariant under $\SU(n)$ transformations). Thus, to exploit the loophole, we must embed a discrete subgroup of $\Uone_\textsc{PQ}$ in a different gauge symmetry, i.e., a gauged flavor symmetry, from the $\SU(n)$ viewpoint. In doing so, we will inevitably encounter difficulties that are closely analogous to those we saw in examples in Sec.~\ref{sec:Zn3}, Sec.~\ref{sec:Uone}, and Sec.~\ref{sec:SUnSUn}, and often worse. For example, in the model of~\cite{Randall:1992ut} (further analyzed in~\cite{Dobrescu:1996jp}), there is an $\SU(n)$ confining group and a further $\SU(m)$ gauge group that is a flavor symmetry from the $\SU(n)$ viewpoint. The naive domain wall number is $nm$, but the lowest dimension gauge invariant operator is baryonic from the $\SU(m)$ viewpoint; effectively, the $\ZZ_m$ center of $\SU(m)$ coincides with a subgroup of $\Uone_\textsc{PQ}$ and is gauged. This reduces the domain wall number from $nm$ to $n$, but does not solve the problem.

While the considerations in this section do not constitute a rigorous no-go theorem, they show that severe difficulties arise in attempts to construct a composite axion model with minimal domain wall number.

\section{Conclusions}
\label{sec:discussion}

We have seen that a wide variety of post-inflation axion models exhibit a fundamental tension: modifying the model to solve the quality problem introduces cosmological problems; modifying the model to solve cosmological problems resurrects the quality problem. The details have varied from model to model, but there are common features. Models that solve the quality problem generally rely on additional gauge symmetries under which the axion transforms. These gauge symmetries can relate different, apparently distinct vacua that are separated by domain walls at the time of the phase transition. The gauge equivalence implies that, in principle, such domain walls can always be destroyed by axion strings. Whether such strings form in the early universe, however, is a detailed dynamical question. A further complication arises from strong experimental constraints on the existence of fractionally charged particles, which require that particles with $\SU(3)_\textsc{C}$ center charge also carry nonzero hypercharge. Adding matter charged under the new gauge symmetries invoked to solve the quality problem, and canceling all gauge anomalies including those involving hypercharge, often requires the introduction of many fields charged under the Standard Model gauge group. These can drive the couplings strong at a low scale $\Lambda$, which appears in the denominator of higher-dimension operators, exacerbating the quality problem.

\subsection{Brief discussion of other models}

Our observations extend to a range of other models incorporating additional physical mechanisms. For example, supersymmetric axion models offer compelling possibilities to solve both the electroweak hierarchy problem and the axion hierarchy problem (i.e., the question of why $f_a \ll M_\mathrm{Pl}$). These problems can be linked, for example through the Kim-Nilles~\cite{Kim:1983dt} mechanism, where a Peccei-Quinn global symmetry forbids a simple $\mu H_u H_d$ term in the superpotential but allows a term of the form $\frac{1}{\Lambda} S^2 H_u H_d$. The phase of $S$ can play the role of the axion, predicting that $\mu \Lambda \sim f_a^2$. To protect the axion quality, a discrete $\ZZ_p$ gauge symmetry~\cite{Babu:2002ic} or $\ZZ_{24}$ gauge R-symmetry~\cite{Babu:2002tx, Lee:2010gv, Lee:2011dya, Baer:2018avn} has been proposed. The arguments of Sec.~\ref{sec:Zn} carry over directly: such models have a domain wall problem in the post-inflation scenario, unless they are extended in some way to explain the cosmological origin of $\ZZ_p$ strings. Any supersymmetric model implementing the Lazarides-Shafi mechanism will differ in two respects from examples we have discussed. On the one hand, holomorphy restricts the set of allowed superpotential terms, potentially ameliorating the quality problem. On the other, supersymmetry increases the number of fields in the theory and thus tends to push Landau poles to lower scales. Perhaps there is a case where the effect of holomorphy is more important, allowing the model to succeed where the models of Sec.~\ref{sec:Uone} and Sec.~\ref{sec:SUnSUn} failed. A search for such models is beyond the scope of this paper.

Recently a novel axion scenario has received some attention, in which a $\ZZ_p$ symmetry acts to permute $p$ copies of the Standard Model that all couple to a common axion~\cite{Hook:2018jle}. This model predicts an exponentially suppressed axion mass, for a given $f_a$, compared to the standard QCD axion, and thus it necessarily predicts a different cosmological history and dark matter abundance than the standard scenario. Nonetheless, one could ask whether such a model admits a viable cosmology with a post-inflation Peccei-Quinn phase transition. This question can only be answered in specific UV completions. A KSVZ-like UV completion of this model, with a complex scalar $S$ charged under $\ZZ_p$ obtaining a VEV and coupling to all $p$ sectors, was argued to solve the axion quality problem~\cite{DiLuzio:2021pxd}. Such a model does not have a viable post-inflation cosmology, because $S$ would be in thermal contact with all $p$ sectors, producing an overwhelming dark radiation problem. A different, composite UV completion discussed in~\cite{DiLuzio:2021pxd} does not solve the quality problem at all.

Our assessment of the quality problem has assumed that operators appear suppressed by a UV cutoff $\Lambda$, below $M_\mathrm{Pl}$ and any Landau poles, with $O(1)$ coefficients. A model like that of Sec.~\ref{sec:SUnSUn}, which solves the domain wall problem and has no fractionally charged relics, could give rise to a viable post-inflation QCD axion if our analysis of the quality problem is flawed. This raises the question: are there axion models in which the coefficients of PQ-violating operators are much smaller than $O(1)$? Within effective field theory, one could simply postulate that there is a Peccei-Quinn symmetry broken only by a very small spurion, which suppresses the coefficients of PQ-violating operators. From our viewpoint, such a model does not solve the Strong CP problem at all; it simply shifts the question of why $|\thetabar|$ is small to the question of why the spurion is small. On the other hand, a UV completion that successfully accounts for a small spurion could solve the quality problem. One approach would be to search for an even larger gauge group that is spontaneously broken at higher energies, but this doesn't add a qualitatively new ingredient compared to our examples and seems unlikely to help. A different physical mechanism is needed. As mentioned in Sec.~\ref{sec:BS_cosmo}, one possibility is that above the Landau pole, one of the theories we have studied matches onto a different EFT that still preserves PQ symmetry up to a larger scale $\Lambda_\textsc{QG}$. Assessing such a scenario would require a more UV-complete model, perhaps a supersymmetric one that undergoes a Seiberg duality at the Landau pole scale. Alternatively, locality in extra dimensions can lead to exponentially small coefficients of higher-dimension operators, if the fields involved are located at sites a distance $L$ apart in extra dimensions and interactions between them are mediated by bulk fields with mass $m \gg 1/L$. This is not immediately useful for a model like that of Sec.~\ref{sec:SUnSUn}, since the PQ-violating operator $\det(Y)$ involves a self-interaction rather than an interaction among fields that can be spatially separated, but there may be models where it succeeds. As already mentioned in Sec.~\ref{sec:intro}, a large class of models that achieves exponential suppression of corrections to the axion potential relies on the axion as a zero mode of an extra-dimensional gauge field~\cite{Witten:1984dg}, but such models have no 4d Peccei-Quinn phase transition and so do not have a post-inflation cosmology. An interesting alternative is the case where a $\Uone$ gauge field obtains a Stueckelberg mass by the 4d Green-Schwarz mechanism~\cite{Green:1984sg, Dine:1987xk}. In this case, the would-be axion is eaten by an anomalous gauge field, but can leave behind an exponentially good approximate global symmetry.\footnote{Ordinarily, when we higgs a $\Uone$ gauge symmetry, there is no good approximate global symmetry left behind. We can insert factors of $\langle \phi \rangle$ to violate charge, and the mass of the gauge field is proportional to $\langle \phi \rangle$, so symmetry breaking effects are large at the higgsing scale. In the Green-Schwarz case, however, charge breaking effects come from factors of $\langle\E^{-T}\rangle$ where the axion $\mathrm{Im}\,T$ shifts under $\Uone$ transformations. The gauge field mass, on the other hand, depends on the second derivative of the kinetic term for $T$ (and is often, parametrically, a UV scale multiplied by a power of $T$). Thus, global symmetry breaking effects can be exponentially small compared to the Stueckelberg mass scale.} The phase of another field charged under the anomalous gauge symmetry can then play the role of the axion; because it is the phase of a 4d field, such models can have an ordinary 4d Peccei-Quinn phase transition. Models with this structure arise in string theory and are often referred to as ``open string axion models,'' although the original example was in the context of the heterotic (closed) string (see, e.g.,~\cite{Barr:1985hk, Kim:1988dd, Svrcek:2006yi, Choi:2011xt, Cicoli:2013cha, Honecker:2013mya, Choi:2014uaa, Buchbinder:2014qca}, or~\cite{Cheng:2001ys} for similar structure in a phenomenological extra dimensional model). These models potentially offer a compelling way out of the problems we have discussed. On the other hand, in the minimal incarnation of this mechanism, the $D$-term potential ensures that the PQ breaking scale is close to the string scale, which would make it difficult to find a model of inflation happening at yet higher energies. It would be interesting to more fully explore the structure of open string axion models that naturally have a separation of scales and the extent to which they can achieve a viable post-inflation axion cosmology.

\subsection{Concluding questions}

We conclude by listing a set of questions raised by this work. To answer some of these questions, we expect that numerical simulations of non-minimal axion models could be useful. Others may be amenable to model-building.

\begin{itemize}

\item What is the string formation dynamics in the model of~\cite{Ardu:2020qmo} with $\SU(N)$ broken to $\SO(N)$ by a symmetric tensor (for odd $N$)? From our discussion in Sec.~\ref{sec:Zn3}, we expect that the strings that form will not be suitable for solving the domain wall problem, but a classical lattice simulation could give a definitive answer. An analytic study of semiclassical string solutions, especially of the relative tension for strings of different winding, could also be instructive.

\item In a scenario with multiple types of strings, such as the model discussed in Sec.~\ref{sec:Uone}, what is the dynamics of string formation? In particular, is the minimal string (on which a single domain wall ends) formed, and in what abundance relative to other strings? How does the answer to this question depend on parameters in the theory (e.g., the relative size of gauge couplings and quartic couplings, or the ratios of different VEVs)?

\item If a string network forms that contains a mix of strings on which single domain walls end and strings on which multiple domain walls end, what happens at the QCD phase transition when the domain wall network forms? What population of minimal strings is sufficient to solve the domain wall problem?

\item Are there viable composite axion models that have domain wall number 1 (possibly invoking the Lazarides-Shafi mechanism)?

\item Are there post-inflation axion models that solve the axion quality problem (and hence the Strong CP problem), have domain wall number 1, and are free of cosmological difficulties? 

\end{itemize}

If the answer to the last question is ``yes,'' then within such models one may be able to predict a target axion dark matter mass, along the lines of recent simulations of axion strings in the post-inflation scenario. However, it may be that these models have non-minimal cosmological dynamics, axion strings of higher tension, or other novelties. Independent of the answer to that question, we believe that the difficulties we have encountered constructing a post-inflation axion scenario that solves the Strong CP problem with a viable cosmology provide a reason to be wary of targeting any specific model or scenario. The QCD axion already provides a well-defined target parameter space across several orders of axion mass. The entire suite of experiments aiming to ``delve deep and search wide''~\cite{Chou:2022luk} is crucial in the search for the axion.

\section*{Acknowledgments}
We have used the {\LaTeX} source of~\cite{Csaki:1996zb} to draw Young tableaux (and also found it a convenient reference for group theory factors). We would like to thank Nima Arkani-Hamed, Roberto Contino, Ben Heidenreich, Anson Hook, Joshua Lin, Maxim Perelstein, Alessandro Podo, Raman Sundrum, and Neal Weiner for helpful discussions. We thank an anonymous referee for useful comments, especially for asking us to clarify the logic of our Landau pole argument. QL is supported by the DOE grant DE-SC0013607, the NSF grant PHY-2210498 and PHY-2207584, and the Simons Foundation. MR is supported in part by the DOE Grant DE-SC0013607. ZS is supported by a fellowship from the MIT Department of Physics. This work was performed in part at the Aspen
Center for Physics, which is supported by National Science Foundation grant PHY-2210452.

\bibliography{ref}
\bibliographystyle{utphys}

\end{document}